\documentclass[12pt]{article}
\usepackage{epsfig}
\usepackage{amsfonts}
\usepackage{latexsym}
\usepackage{amsmath,amssymb}
\usepackage{mathrsfs}
\usepackage{hyperref}
\usepackage{setspace}
\usepackage{color}
\usepackage{bm}
\usepackage{slashed}
\usepackage{braket}
\numberwithin{equation}{section}

\newcommand{\tf}{t_{\rm f}}

\begin{document}

\begin{titlepage}
\unitlength = 1mm
\begin{flushright}
KOBE-COSMO-21-19
\end{flushright}

\vskip 1cm
\begin{center}

{\large {\textsc{\textbf{ Squeezed quantum states of graviton and axion in the universe}}}}

\vspace{1.8cm}
Sugumi Kanno$^*$ and Jiro Soda$^{\flat}$

\vspace{1cm}

\shortstack[l]
{\it $^*$ Department of Physics, Kyushu University, 744 Motooka, Nishi-ku, Fukuoka 819-0395, Japan \\ 
\it $^\flat$ Department of Physics, Kobe University, Kobe 657-8501, Japan
}

\vskip 1.5cm

\begin{abstract}
\baselineskip=6mm
Particle production during cosmic expansion
can be interpreted as a two-mode squeezing process of quantum states. The two-mode squeezed states consist of an infinite number of entangled particles and then enhance the nonclassicality of particles. There are two methods for estimating the degree of squeezing.
One is to employ instantaneous vacuum states, and the other is to adopt adiabatic vacuum states. We analytically study the squeezing
process of gravitons and axions
by using the two methods. We first consider a cosmological model of inflation followed by instantaneous reheating leading to a radiation dominated era. In the case of gravitons, we find no qualitatively difference in the squeezing between the two methods. However, for the axions, it turns out that the squeezing in the instantaneous vacuum increases as the mass increases while the squeezing decreases in the adiabatic vacuum as the mass increases. We then study the effect of non-instantaneous reheating on the the squeezing and show that the squeezing is enhanced compared with the instantaneous reheating.
We also illustrate how the squeezed states 
enhance the violation of Bell inequality and quantum noise of gravitons and axions.
\end{abstract}

\vspace{1.0cm}

\end{center}
\end{titlepage}

\pagestyle{plain}
\setcounter{page}{1}
\newcounter{bean}
\baselineskip18pt

\setcounter{tocdepth}{2}

\tableofcontents

\section{ Introduction}

One of the cornerstones of the  inflationary cosmology is that large scale structure of the universe stems from quantum fluctuations. Due to the rapid cosmological expansion, the quantum fluctuations are squeezed as they exit the horizon~\cite{Grishchuk:1989ss,Grishchuk:1990bj,Albrecht:1992kf,Polarski:1995jg}. 
From the point of view of quantum information theory, this squeezing process is interpreted as a process of creation of  entangled particles. However, no compelling observational evidence for the nonclassicality stemming from the inflation has yet been found. It is expected that primordial gravitational waves (PGWs) and axions \cite{Peccei:1977hh,Weinberg:1977ma,Wilczek:1977pj,Kim:1979if,Shifman:1979if,Dine:1981rt} may help us to find a signature of quantum mechanical origin of the universe because they arise out of quantum fluctuations directly and their interactation with matter is so weak that their quantum nature may survive until today. If nonclassical PGWs or axions are detected, it would be an evidence of the quantum mechanical origin of the universe and at the same time, it would imply a discovery of gravitons and axions. 

In~\cite{DYSON:2013jra,Rothman:2006fp}, they demonstrated that no conceivable experiment in the real universe can detect a single graviton directly. However, there appears a possibility to detect gravitons indirectly through the noise of gravitons on the gravitational interferometers~\cite{Parikh:2020nrd,Kanno:2020usf,Parikh:2020kfh,Parikh:2020fhy} or by measuring the decoherence time of an object caused by the noise of gravitons~\cite{Kanno:2021gpt}. The point in the literature is that gravitons are in a squeezed state currently. 


Recently, axions proposed for resolving the strong CP problem in QCD have been intensively studied as a dark matter candidate~\cite{Preskill:1982cy,Abbott:1982af,Dine:1982ah,Ipser:1983mw}.
Since axions in pre-inflationary scenario also get squeezed during inflation, the nonclassicality of the axions could survive until today in dark matter. In~\cite{Kanno:2021akn}, an idea to detect axions by havesting quantum coherence of axions is proposed. 


Thus, it is worth studying the squeezing process of quantum states of gravitons and axions in order to observe the quantum mechanical origin of the universe. There are two methods for estimating the degree of squeezing owing to a difficulty in specifying the vacuum state in a time dependent spacetime. One is to employ instantaneous vacuum states~\cite{Grishchuk:1989ss,Grishchuk:1990bj,Albrecht:1992kf,Polarski:1995jg}, and the other is to adopt adiabatic vacuum states~\cite{BirrellDavies}. The instantaneous vacuum state we call in this paper is defined at every moment from inflation till radiation dominated phases while the adiabatic vacuum state is defined in the remote past during inflation phase and in the remote future in the radiation dominated phase.
Recently, the squeezing parameters of axions are calculated numerically in the instantaneous vacuum state~\cite{Kuss:2021gig}.
However, the standard approach to discuss particle creation in curved space is to use the adiabatic vacuum. Hence, we need to clarify whether the squeezing process can be described identically even in the different vacuum states. 
In this paper, we derive an analytic formula for the squeezing parameters of axions (massive scalar fields) as well as gravitons.
We also investigate the effect of reheating phase
on the squeezing of gravitons and axions.

It should be noted that the result of  \cite{Polarski:1995jg} is often erroneously mentioned.
The commutator of canonical variables never vanishes even in the squeezing limit during inflation. 
Hence, squeezing of gravitons and axions
never implies classicality.
To emphasize the nonclassicality of squeezed state, we  demonstrate that Bell inequalities~\cite{Bell:1964kc,Maldacena:2015bha,Kanno:2017dci,Choudhury:2017bou,Ando:2020kdz} can be  a measure of entanglement of squeezed state, and how the quantum noises of gravitons and axions
are enhanced in squeezed states.

The paper is organized as follows. 
In section 2, we set up our cosmological model.
In section 3, we calculate the squeezing parameters of gravitons in the instantaneous and the adiabatic vacuum respectively. In section 4, we compute the squeezing parameters of axions. In particular, we derive an analytic formula for the Bogiliubov coefficients in the instantaneous and the adiabatic vacuum respectively. We show that the adiabatic vacuum is suitable for describing the squeezing of axions. In section 5, we take into account the effects of reheating phase on the squeezing.
In section 6, we discuss implications of squeezed states 
by using some measures of nonclassicality.  
Finally, our results are summarized and discussed in Section 7.

\section{Cosmological set up}

In this paper, for simplicity, we focus on primordial fluctuations that re-enter the horizon in the radiation dominated era. Namely, we do not consider primordial fluctuations with the wavelength larger than 10 Mpc which are affected by a transition from a radiation dominated phase to a matter dominated phase, and by a subsequent transition to dark-energy dominated phase.

We first consider instantaneous reheating after inflation approximated by the de Sitter phase leading to a radiation dominated phase. In other words, the scale factor evolves as follows
\begin{eqnarray}
 a (\eta )=\left\{
\begin{array}{l}
\vspace{0.2cm}
 \frac{1}{-H  (  \eta -2 \eta_1 )}  \hspace{1.2cm}   {\rm for}  \quad   \eta   <  \eta_1\,,\\
\frac{\eta}{H\eta^2_1} \hspace{2.2cm}     {\rm for} \quad     \eta_1  <  \eta   \ ,
\end{array}
\right.             
\label{history}
\end{eqnarray}
where $\eta_1$ is the conformal time of reheating. The scale factor is smoothly connected up to the first order of derivative at the reheating. 
Here, $H$ is the Hubble parameter during inflation. The effect of a reheating phase will be discussed later in Section~\ref{5}.

As we mentioned, primordial fluctuations
are generated during inflation.
Hence, the highest wavenumber of the
fluctuations exits the horizon at the reheating time. Namely,
the highest frequency of the primordial fluctuations generated during inflation is determined by the time $\eta_1$.
At the time of reheating, the scale factor is given by 
\begin{eqnarray}
a(\eta_1)\equiv a_1 = \frac{1}{H\eta_1} \ .
\end{eqnarray}
The comoving wavenumber of the fluctuations generated at the $\eta_1$ is given by $k_1=2\pi f_1$ where $f_1$ is the cutoff frequency and there are no fluctuations generated during inflation with frequency higher than the $f_1$. We suppose that the fluctuations re-enter the horizon with the wavenumber which satisfies $H^{-1}=a_1/k_1$. Hence, the cutoff frequency is expressed by $f_1=Ha_1/(2\pi)$.
Since we consider instantaneous reheating leading to a radiation dominated era, we have the following relation 
\begin{eqnarray}
\frac{a_1}{a_0} = \frac{T_0}{T_{\rm reh}} \,,
\end{eqnarray}
where $T_0$ and $T_{\rm reh}$ are the observed temperature and the reheating temperature.  The primordial fluctuations undergo redshift and the cutoff frequency when observed is estimated as 
\begin{eqnarray}
f_1\frac{a_1}{a_0} = \frac{1}{2\pi}\frac{HT_0}{ T_{\rm reh}} 
=  10^{10}\, {\rm Hz} \left( \frac{T_0}{2.7{\rm K}}\right)
\left( \frac{10^{14}\,{\rm GeV}}{T_{\rm reh}}\right)
\left( \frac{H}{10^{14}\,{\rm GeV}}\right)\ .
\end{eqnarray}
Note that the result is almost the same even if we take into account the redshift of matter dominated era. The main ambiguity of the cutoff frequency comes from the reheating
temperature.

With the above set up, we investigate the squeezing process of gravitons and axions separately focusing on the dependence of the vacuum state in the following.
For the axions, we also focus on the mass dependence on the squeezing which clarifies the difference between gravitons and axions.

\section{Squeezed quantum states of gravitons}
\label{3}

In this section, we study the gravitational particle production of gravitons. To discuss the particle production, one needs to define the vacuum state that seems to be the most natural in a time dependent spacetime. In cosmology, there are two formalisms that apply instantaneous vacuum states and adiabatic vacuum states. We first apply the instantaneous vacuum state for calculating Bogoliubov coefficients developed by ~\cite{Grishchuk:1990bj,Albrecht:1992kf,Polarski:1995jg} in Section~{\ref{3.1}}. Then we apply the adiabatic vacuum state by~\cite{BirrellDavies} in Section~{\ref{3.2}}.
We compare the results between two formalisms  in Section~{\ref{3.3}}.

Let us consider the metric perturbations $h_{ij}(\eta,x^i)$
\begin{eqnarray}
  ds^2   =  a^2 (\eta )  \left[ -d\eta^2 
  +   \left(\delta_{ij} +h_{ij}(\eta,x^i) \right)dx^i dx^j     \right]  \ ,
\end{eqnarray}
where $\eta$ and $x^i$ are the conformal time and spatial coordinates, $a(\eta)$ is the scale factor, $\delta_{ij}$ is the Kronecker delta and $h_{ij}$ satisfies the transverse traceless conditions $h_{ij,j}=h_{ii}=0$. The indices $(i,j)$  run from 1 to 3.
The gravitational fields in the homogeneous and isotropic universe are equivalent to massless scalar fields.
In fact, if we expand the gravitational field in terms of its corresponding Fourier modes
\begin{eqnarray}
  h_{ij} ({\bf x} ,\eta) = 
  \frac{2}{M_p }\sum_{P}\frac{1}{\left(2\pi\right)^{3/2}}\int d^3 {\bf k} 
 \ h_P ({\bf k} ,\eta)  e^P_{ij}({\bf k})
 e^{i{\bf k}\cdot{\bf x}}
     \ ,
\end{eqnarray}
where $e^P_{ij}$ is the polarization tensor normalized as $e^{P*}_{ij}({\bf k})e^Q_{ij}({\bf k})=\delta^{PQ}$ and the index $P$ denotes the polarization modes, for instance, for circular polarization modes $P=\pm$ and for linear polarization modes $A=+,\times$. The  action in the Fourier modes becomes identical to that of massless scalar field such as
\begin{eqnarray}
  S = \int d^3 {\bf x }\,d\eta\,L
  =  \frac{1}{2}\int d^3{\bf k}\,d\,\eta\,a^2   \left[  \left|h'_{\bf k}\right|^2 
  -k^2 \left| h_{\bf k}\right|^2\right]   
     \,,
\end{eqnarray}
We used $k=|{\bf k}|$. 
Hereafter, we omit the polarization index $P$ for simplicity unless there may be any confusion and write $h_P({\bf k},\eta)\equiv h_{\bf k}(\eta)$ in the following.

\subsection{Squeezing in instantaneous vacuum states}
\label{3.1}

In field theory, we define mode functions so that we can define Poincare invariant vacuum.
Then properly normalized mode functions lead to creation and annihilation  operators. 
However, in curved spacetime, there is no Poincare invariance. 
Hence, we cannot specify a unique mode function. 
Instead, we can directly define creation and annihilation operators
in terms of canonical variables. We call the vacuum defined by this annihilation operator the instantaneous vacuum as we will explain below.

It is convenient to define a new variable $y_{\bf k}(\eta)=a(\eta)h_{\bf k}(\eta)$. The Lagrangian becomes
\begin{eqnarray}
  L= \int d^3{\bf k}\,\frac{1}{2}  \left[  y_{\bf k}^{\prime }\, y_{-\bf k}^{\prime } 
  -   k^2 y_{\bf k}\, y_{-\bf k} -    \frac{a'}{a}  y_{\bf k}\, y_{-\bf k}'  
  -    \frac{a'}{a}  y_{-\bf k}\, y_{\bf k}'  
  + \left( \frac{a'}{a} \right)^2 y_{\bf k}\, y_{-\bf k}\right] \,.
\end{eqnarray}
where primes denote derivatives with respect to the conformal time. The conjugate momentum is defined by
\begin{eqnarray}
  p_{\bf k} = \frac{\partial L}{\partial y_{-\bf k}'}  
  =  y_{\bf k}' -   \frac{a'}{a}  y_{\bf k}  \ .
  \label{conjugate}
\end{eqnarray}
Then the Hamiltonian in Fourier space is given by
\begin{eqnarray}
  \int d^3{\bf k}\,H_{\bf k} = \frac{1}{2}  \int d^3{\bf k}\left[  p_{\bf k}\,p_{\bf -k}   + k^2 y_{\bf k}\,y_{\bf -k}   +    \frac{a'}{a}\left( y_{\bf k}\,p_{\bf -k} + p_{\bf k}\, y_{\bf -k} \right)\right] \ ,
\end{eqnarray}
where $H_k$ is a Hamiltonian density in $k$ space. 
The equations of motion are derived from the Hamiltonian in the form,
\begin{eqnarray}
y_{\bf k}' &=& \frac{\partial H_k}{\partial p_{-\bf k}}  =  p_{\bf k} +  \frac{a'}{a}  y_{\bf k}\,,
\label{eom-y1}\\
  p_{\bf k}' &=& -\frac{\partial H_k}{\partial y_{-\bf k}}  = -k^2  y_{\bf k}  -   \frac{a'}{a}  p_{\bf k}\,.
\label{eom-p1}
\end{eqnarray}

For later convenience, it is useful to derive the second order differential equations for the canonical variables by using the above Eqs.~(\ref{eom-y1}) and (\ref{eom-p1}) :
\begin{eqnarray}
 &&y_{\bf k}'' +   \left(   k^2  -    \frac{a''}{a}  \right) y_{\bf k}   = 0\,,
\label{eom-y2}\\
 && p_{\bf k}''  +   \left(   k^2  +   a\left(\frac{a^\prime}{a^2}\right)^\prime\,  \right) p_{\bf k}   = 0
\label{eom-p2} \ .
\end{eqnarray}
Note that $p_{\bf k}$ and $y_{\bf k}$ are related to each other through the relation Eq.~(\ref{conjugate}).

Now we quantize the position $y_{\bf k}$ and the momentum $p_{\bf k}$. Then the position operator $\hat{y}_{\bf k}$ and the momentum operator $\hat{p}_{\bf k}$ is represented by creation and annihilation operators $\hat{a}_{\bf k}$, $\hat{a}^\dag_{-\bf k}$ such as
\begin{eqnarray}
\hat{y}_{\bf k} (\eta )&=& \frac{1}{\sqrt{ 2k}}  \left(   \hat{a}_{\bf k} (\eta) +  \hat{a}^\dagger_{-\bf k} (\eta) \right)\,, 
\label{operator-y}\\
\hat{p}_{\bf k} (\eta )&=& -i \sqrt{ \frac{k}{2}}  \left(   \hat{a}_{\bf k} (\eta) -  \hat{a}^\dagger_{-\bf k} (\eta) \right)
\label{operator-p}\,.
\end{eqnarray}
The commutation relation $[ \hat{a}_{\bf k} (\eta), \hat{a}^\dagger_{-\bf k'}(\eta)]=\delta ({\bf k}+{\bf k'})$
guarantees the canonical commutation relation $[ y_{\bf k}(\eta),p_{\bf k'}(\eta)]=i\delta_{{\bf k},{\bf k'}}$.
It is useful to write the annihilation operator by using the canonical variables as
\begin{eqnarray}
  \hat{a}_{\bf k} (\eta) =
  \sqrt{ \frac{k}{2}}\,\hat{y}_{\bf k} (\eta )   + \frac{i}{\sqrt{ 2k}}\,\hat{p}_{\bf k} (\eta )\,.
  \label{def_ann}
\end{eqnarray}
Note that the annihilation operator becomes time dependent through the time dependence of canonical variables. Hence, the vacuum defined by $\hat{a}_{\bf k}(\eta)|0\rangle=0$ is time dependent as well and the vacuum in this formalism turns out to be defined at every moment. We call this vacuum state instantaneous vacuum state.
In the literature~\cite{Grishchuk:1990bj,Albrecht:1992kf,Polarski:1995jg}, the time evolution of the creation and annihilation operators is described by the Bogoliubov transformations such as
\begin{eqnarray}
  \hat{a}_{\bf k} (\eta) 
  =   \alpha_k (\eta )\,\hat{a}_{\bf k} (\eta_0  )  + \beta_k (\eta )\,\hat{a}^\dagger_{-\bf k} (\eta_0) \ ,
\end{eqnarray}
where $\eta_0$ is an initial time. The Bogoliubov coefficients $\alpha_k$ and $\beta_k$ should satisfy the normalization condition $|\alpha_k (\eta )|^2 -|\beta_k (\eta )|^2 =1$ so that the commutation relation $[ \hat{a}_{\bf k} (\eta), \hat{a}^\dagger_{-\bf k'}(\eta)]=\delta ({\bf k}+{\bf k'})$ holds.
The most standard parametrization of the Bogoliubov coefficients is given by
\begin{eqnarray}
  \alpha_k (\eta )   &=&  e^{-i\theta_k(\eta ) }  \cosh r_k (\eta) \,,   \\
  \beta_k (\eta )  &=&  e^{i\theta_k(\eta ) +2i \varphi_k (\eta ) }  \sinh r_k (\eta) 
 \ ,
\end{eqnarray}
where $r_k , \theta_k , \varphi_k$ are called the squeezing parameter and the phases, respectively. These parameters are determined by using the above formulae after the solution $\hat{y}_{\bf k} (\eta)$ is obtained. Then the Bogoliubov coefficients are calculated. Note that the effective phase of the Bogoliubov coefficients is given by
\begin{eqnarray}
\theta_k(\eta )+\varphi_k (\eta )=\frac{1}{2}\arccos{\left({\rm Re}\left[\frac{|\alpha_k(\eta)|}{\alpha_k(\eta)}\frac{\beta_k(\eta)}{|\beta_k(\eta)|}\right]\right)}\,.
\label{phase-instantaneous}
\end{eqnarray}
Next, we solve the equations of motion 
for the modes of gravitons, Eqs.~(\ref{eom-y2}) and (\ref{eom-p2}), in each phase of the universe, separately.

\subsubsection{Evolution in de Sitter  universe}
\label{3.1.1}

Let us start with the de Sitter phase.
Then Eqs.~(\ref{eom-y2}) and (\ref{eom-p2}) are written as
\begin{eqnarray}
 &&\hat{y}_{\bf k}'' +   \left(   k^2  -    \frac{2}{(\eta-2\eta_1)^2}  \right) \hat{y}_{\bf k}   = 0   \ , 
 \label{eom-y-ds}\\
 && \hat{p}_{\bf k}''  +     k^2   \hat{p}_{\bf k}   = 0 
 \label{eom-p-ds}\ .
\end{eqnarray}
The solution of Eq.~(\ref{eom-y-ds}) is
\begin{eqnarray}
\hat{y}_{\bf k}  =   \frac{1}{\sqrt{ 2k}}\left(1-\frac{i}{k(\eta-2\eta_1)}  \right) e^{  - ik(\eta-2\eta_1)} \hat{A} 
+  \frac{1}{\sqrt{ 2k}} \left(1+  \frac{i}{k(\eta-2\eta_1)}  \right)
e^{ ik(\eta-2\eta_1)} \hat{A}^\dagger   \ ,
\end{eqnarray}
where $A$ and its conjugate $A^\dagger$ are constant operators of integration. Note that we chose the properly normalized positive frequency mode in the remote past where the Minkowski vacuum is defined.
The solution of Eq.~(\ref{eom-p-ds}) is the conjugate momentum expressed as
\begin{eqnarray}
\hat{p}_{\bf k}  =  -i \sqrt{ \frac{k}{2}}\,  e^{  - ik(\eta-2\eta_1)} \hat{A} +  i \sqrt{ \frac{k}{2}}\, e^{ ik(\eta-2\eta_1)} \hat{A}^\dagger \ .
\end{eqnarray}
By using Eq.~(\ref{def_ann}), we obtain the time evolution of the anihilation operator in the Heisenberg picture of the form
\begin{eqnarray}
  \hat{a}_{\bf k} (\eta) &=&
  \sqrt{ \frac{k}{2}}\,\hat{y}_{\bf k} (\eta )   + \frac{i}{\sqrt{ 2k}}\,\hat{p}_{\bf k} (\eta )   \nonumber \\
  &=&  \left(1-\frac{i}{2 k (\eta-2\eta_1) }  \right) e^{  - ik (\eta-2\eta_1)} \hat{A} 
               +  \frac{i}{2 k (\eta-2\eta_1)}  e^{ ik (\eta-2\eta_1)}\hat{A}^\dagger
               \label{annihilation-ds}\,.
\end{eqnarray}
In order to find $\hat{A}$ and $\hat{A}^\dag$, we give the initial condition at time $\eta=\eta_0$ such as
\begin{eqnarray}
  \hat{a}_{\bf k} (\eta_0)   =  \left(1-\frac{i}{2 k (\eta_0 -2\eta_1) }  \right) e^{  - ik (\eta_0 -2\eta_1)} \hat{A} 
               +  \frac{i}{2 k (\eta_0 -2\eta_1)}  e^{ ik (\eta_0 -2\eta_1)}\hat{A}^\dagger   
               \ .
\end{eqnarray}
Combining with the Hermitian conjugate of the above equation, we find $\hat{A}$ and $\hat{A}^\dag$ in terms of $\hat{a}_{\bf k} (\eta_0)$ and $\hat{a}^\dag_{-\bf k} (\eta_0)$.
Substituting them back into Eq.~(\ref{annihilation-ds}), we obtain the Bogoiubov transformation in de Sitter phase
\begin{eqnarray}
\hat{a}_{\bf k} (\eta)=\alpha^{\rm I}_k(\eta)\,\hat{a}_{\bf k} (\eta_0)+\beta^{\rm I}_k(\eta)\,\hat{a}^\dagger_{-{\bf k}}(\eta_0)\,,
\label{bogoiubov-ds}
\end{eqnarray}
where the Bogoliubov coefficients are
\begin{eqnarray}
\alpha^{\rm I}_k(\eta)=
\left(1-\frac{i}{2 k (\eta -2\eta_1) } \right) \left(1+ \frac{i}{2 k (\eta_0 -2\eta_1) }  \right)
                    e^{ -ik (\eta-\eta_0)}
                  -  \frac{1}{4k^2 (\eta -2\eta_1) (\eta_0 -2\eta_1)} e^{ ik (\eta- \eta_0)}\,,
                  \hspace{9mm}\\
\beta^{\rm I}_k(\eta)=
\frac{i}{2 k (\eta -2\eta_1) }  \left(1- \frac{i}{2 k (\eta_0 -2\eta_1) }  \right)
                    e^{ ik (\eta-\eta_0)}
                  - \frac{i}{2 k (\eta_0 -2\eta_1) }  \left(1- \frac{i}{2 k (\eta -2\eta_1) }  \right)
                  e^{- ik (\eta- \eta_0)}\,.\hspace{5mm}
\end{eqnarray}

\subsubsection{Evolution in radiation dominant universe}
\label{3.1.2}

After instanteneous reheating, the universe enters into the radiation dominated phase. Then Eqs.~(\ref{eom-y2}) and (\ref{eom-p2}) are
\begin{eqnarray}
 &&\hat{y}_{\bf k}'' +    k^2\, \hat{y}_{\bf k}   = 0  
 \label{eom-y-rad}\ , \\
 && \hat{p}_{\bf k}''  +   \left(   k^2  - \frac{2}{\eta^2}   \right) \hat{p}_{\bf k}   = 0 
 \label{eom-p-rad}\ .
\end{eqnarray}
Note that comparing the above with Eqs.~(\ref{eom-y-ds}) and (\ref{eom-p-ds}), we see that the roles of $y_{\bf k}$ and $p_{\bf k}$ are interchanged between de Sitter and radiation dominated phases.

The solution of Eq.~(\ref{eom-y-rad}) is 
\begin{eqnarray}
\hat{y}_{\bf k}  = 
  \frac{1}{\sqrt{ 2k}} e^{  - ik\eta}\,\hat{B} +  \frac{1}{\sqrt{ 2k}} e^{ ik\eta}\,\hat{B}^\dagger  \,,
\end{eqnarray}
where $\hat{B}$ and $\hat{B}^\dag$ are constant operators of integration. Note that we chose $1/\sqrt{2k}\,e^{\pm ik\eta}$ simply as independent solutions.
 In this case, the conjugate momentum is given by
\begin{eqnarray}
\hat{p}_{\bf k}  = -i \sqrt{ \frac{k}{2}}   \left(1-\frac{i}{k\eta}  \right) e^{  - ik\eta}\,\hat{B} 
+ i \sqrt{ \frac{k}{2}}  \left(1+  \frac{i}{k\eta}  \right) e^{ ik\eta}\,\hat{B}^\dagger  \ .
\end{eqnarray}
Thus, the annihilation operator is obtained by using Eq.~(\ref{def_ann}) such as
\begin{eqnarray}
  \hat{a}_{\bf k} (\eta) 
  =  \left(1-\frac{i}{2 k\eta}  \right) e^{  - ik\eta}\,\hat{B} 
               -  \frac{i}{2 k\eta}  e^{ ik\eta}\,\hat{B}^\dagger   \,.
\end{eqnarray}
The initial condition of $\hat{a}_{\bf k}$ is given at the reheating time $\eta_1$ and $\hat{B}$ and $\hat{B}^\dag$ are determined in the same way as explained in the de Sitter phase.
Then we find the Bogoliubov transformation in radiation dominated phase
\begin{eqnarray}
\hat{a}_{\bf k}(\eta)= \alpha^{\rm R}_k (\eta)\,\hat{a}_{\bf k} (\eta_1)+\beta^{\rm R}_k(\eta)\,\hat{a}^\dagger_{-{\bf k}} (\eta_1) 
\label{bogoliubov-rad}\ ,
\end{eqnarray}
where the Bogoliubov coefficients are
\begin{eqnarray}
\alpha^{\rm R}_k (\eta)&=&\left(1+\frac{1}{4 k^2 \eta  \eta_1 }
-\frac{i}{2k\eta}+\frac{i}{2k\eta_1} \right)  e^{  - ik (\eta -\eta_1)} 
               -  \frac{1}{4 k^2 \eta \eta_1}  e^{ ik (\eta - \eta_1)}\,,\\
\beta^{\rm R}_k(\eta)&=&\frac{i}{2 k \eta_1} \left(1-\frac{i}{2 k \eta }  \right) e^{  - ik (\eta - \eta_1)} 
               -  \frac{i}{2 k \eta}  \left(1-\frac{i}{2 k \eta_1 }  \right) 
               e^{ ik (\eta -\eta_1)} \ .
\end{eqnarray}

\subsubsection{Squeezing parameters}
Since we obtained the Bogoliubov coefficients in the de Sitter and the radiation dominated phases respectively, we can find the total Bogoliubov coefficients from the initial time in de Sitter phase to the arbitrary time in radiation dominated phase by combining the previous results. That is, we plug Eq.~(\ref{bogoiubov-ds}) and its complex conjugation at $\eta=\eta_1$:
\begin{eqnarray}
\hat{a}_{\bf k} (\eta_1)&=&\alpha^{\rm I}_k(\eta_1)\,\hat{a}_{\bf k} (\eta_0)+\beta^{\rm I}_k(\eta_1)\,\hat{a}^\dagger_{-{\bf k}} (\eta_0)\,,\\
\hat{a}_{-{\bf k}}^\dagger (\eta_1)&=&\alpha^{{\rm I}*}_k(\eta_1)\,\hat{a}^\dagger_{-{\bf k}} (\eta_0)+\beta^{{\rm I}*}_k(\eta_1)\,\hat{a}_{\bf k} (\eta_0)  \ .
\end{eqnarray}
into Eq.~(\ref{bogoliubov-rad}), then we have 
\begin{eqnarray}
\hat{a}_{\bf k}(\eta)=\alpha_k(\eta)\hat{a}_{\bf k}(\eta_0)+\beta_k(\eta)\hat{a}^\dagger_{-{\bf k}}(\eta_0) \ ,
\end{eqnarray}
where we can read off the total Bogoliubov coefficients
\begin{eqnarray}
\alpha_k (\eta)&=&\alpha^{\rm R}_k(\eta)\alpha^{\rm I}_k(\eta_1)+\beta^{\rm R}_k(\eta)\beta^{{\rm I}*}_k(\eta_1)\,,\\
\beta_k(\eta)&=&\alpha^{\rm R}_k(\eta)\beta^{\rm I}_k(\eta_1)+\beta^{\rm R}_k(\eta)\alpha^{{\rm I}*}_k(\eta_1) \ .
\end{eqnarray}
If we consider the limits of remote past $-k\eta_0\gg 1$ and remote future 
$ k^2 \eta_1\eta\gg 1$,
we find
\begin{eqnarray}
  \alpha_k (\eta)  
   &=&  \left(  1+\frac{i}{k \eta_1 } 
     -  \frac{1}{2 k^2 \eta_1^2 } \right)
                    e^{ -ik (\eta-\eta_0)}
\label{alpha-instantaneous}\,,\\
  \beta_k (\eta)  
   &=&    \frac{1}{2 k^ 2 \eta_1^2 }\, 
       e^{  - ik (\eta - 2\eta_1 +\eta_0)} 
\label{beta-instantaneous}\ .
\end{eqnarray}

\begin{figure}[t]
\begin{center}
\vspace{-2cm}
\hspace{-1.2cm}
\begin{minipage}{8cm}
\includegraphics[height=6.5cm]{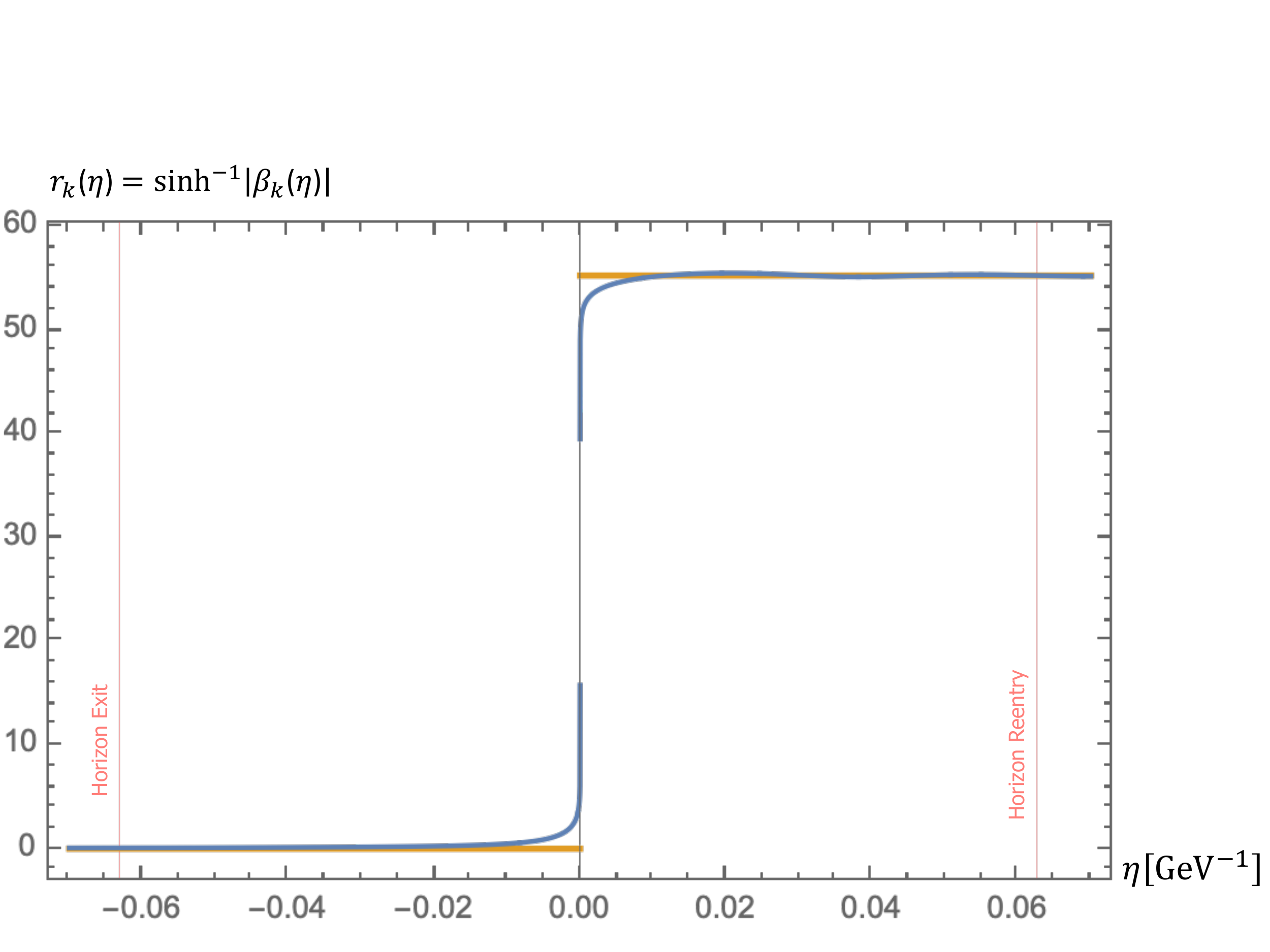}\centering
\end{minipage}
\begin{minipage}{8cm}
\hspace{0.4cm}
\includegraphics[height=6.5cm]{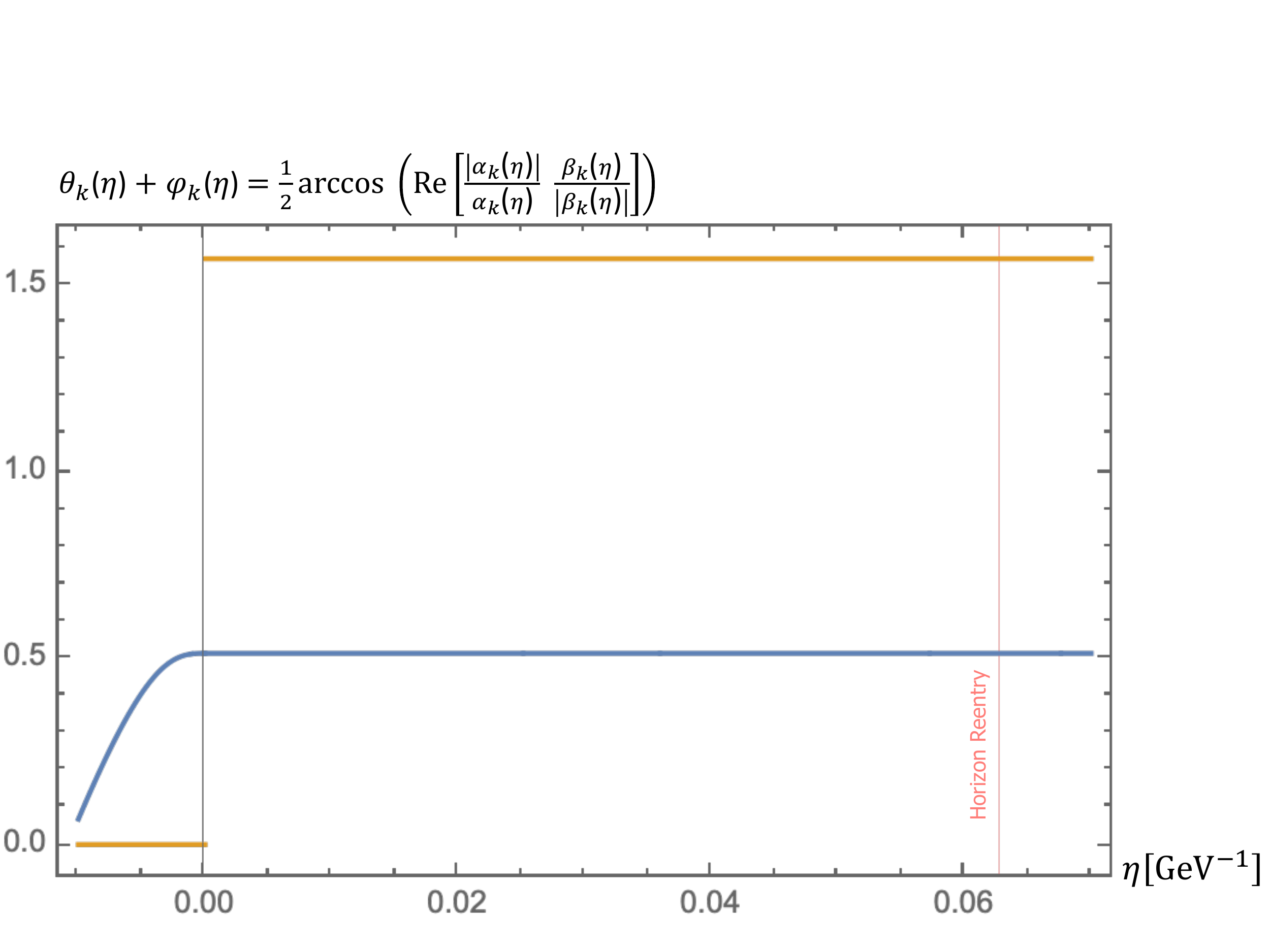}
\end{minipage}
\caption{The left panel shows the time evolution of the squeezing parameter $r_k$ for $k=10^2$ GeV, $\eta_1=10^{-14}$ ${\rm GeV}^{-1}$ and $H=10^{14}$ GeV by using instantaneous vacuum state (blue) and adiabatic vacuum state (orange). The vertical red lines show the time of horizon exit $-2\pi/k$ and re-entry $2\pi/k$.
The right panel shows the time evolution of the effective phase $\theta_k+\varphi_k$ with the same parametrization.}
\label{fig1}
\end{center}
\end{figure}

The squeezing parameters are written as
\begin{eqnarray}
  r_k (\eta)  
   &=&  \sinh^{-1} |\beta_k (\eta) |   \,,  \label{r-instantaneous}\\
  \theta_k (\eta) 
   &=&   \arccos\left(\rm{Re}\left[\frac{\alpha_k (\eta)}{|\alpha_k (\eta) |}\right]\right)
  \,,\\
     \varphi_k (\eta)  
   &=& \frac{1}{2} \arccos\left(\rm{Re}\left[ \frac{\alpha_k (\eta) \beta_k (\eta)}{|\alpha_k (\eta)  \beta_k (\eta)|}\right]\right)\ .
\end{eqnarray}
We plotted in blue the results of the squeezing parameter Eq.~($\ref{r-instantaneous}$) and the effective phase Eq.~($\ref{phase-instantaneous}$) in the range $-k\,\eta_0\gg 1$, $k^2 \eta_1\eta\lesssim1$, respectively in the left and right panel in Figure~{\ref{fig1}}.

\subsection{Squeezing in adiabatic vacuum states}
\label{3.2}

In this subsection, we assume the vacuum state can be specified by the adiabatic vacuum which behaves in the same way as the Minkowski vacuum state in the remote past in de Sitter phase and in the remote future in radiation dominated phase, respectively~\cite{BirrellDavies}. Thus, the Bogoliubov coefficients
between two the different vacua are time independent.  

In de Sitter phase, the positive frequency mode of Eq.~(\ref{eom-y-ds}) satisfying the normalization
$i(u_k^*u_k'-u_ku^{*\prime}_k)=1$ is given by
\begin{eqnarray}
u_k  =   \frac{1}{\sqrt{ 2k}}\left(1-\frac{i}{k(\eta-2\eta_1)}  \right) e^{  - ik (\eta-2\eta_1)}\,, \qquad{\rm for}\quad \eta<\eta_1\,.
\label{uk}
\end{eqnarray}
The positive frequency mode of Eq.~(\ref{eom-y-rad}) in the radiation dominated phase is given by
\begin{eqnarray}
v_k  =   \frac{1}{\sqrt{ 2k}} e^{  - ik\eta}\,, \qquad{\rm for}\quad\eta>\eta_1\,.
\end{eqnarray}
This mode function and its complex conjugate $v^*_{-k}$ constitute a different basis in the solution space of Eq.~(\ref{eom-y2}).
Hence, the mode function $u_k$ in the radiation dominated phase is written by $v_k$ such as
\begin{eqnarray}
  u_k (\eta) 
  =   \alpha_k  v_k (\eta  )  + \beta_k  v^*_{-k} (\eta) \ ,\qquad{\rm for}\quad\eta>\eta_1\,,
\end{eqnarray}
where the Bogoliubov coefficients $\alpha_k$, $\beta_k$ are constant in contrast to Eqs.~(\ref{alpha-instantaneous}), (\ref{beta-instantaneous}) because they are determined by the continuity relations of the mode functions $u_k$ and $v_k$ at the reheating time $\eta_1$:
\begin{eqnarray}
  u_k (\eta_1) 
  &=&   \alpha_k  v_k (\eta_1  )  + \beta_k  v^*_{-k} (\eta_1)  \ ,  
  \label{bogoliubov1}\\
   u'_k (\eta_1) 
  &=&   \alpha_k  v'_k (\eta_1  )  + \beta_k  v^{*\prime}_{-k} (\eta_1) 
  \label{bogoliubov2}\ .
\end{eqnarray}
The Bogoliubove coefficients are given by
\begin{eqnarray}
  \alpha_k    &=&  \left(1+\frac{i}{k  \eta_1}  - \frac{1}{2 k^2  \eta_1^2} \right)   e^{2 i k  \eta_1  } \,,     \\
  \beta_k   &=&  \frac{1}{2k^2\eta_1^2}     \ .  
  \label{bogoliubov-beta}
\end{eqnarray}
Note that the phase of the Bogoliubov coefficients is different from Eqs.~(\ref{alpha-instantaneous}) and (\ref{beta-instantaneous}).
The squeezing parameter Eq.~(\ref{r-instantaneous}) and the effective phase Eq.~(\ref{phase-instantaneous}) in this formalism are written as
\begin{eqnarray}
r_k &=& \sinh^{-1} |\beta_k |
\,, \\
\theta_k+\varphi_k&=&\frac{1}{2}\arccos{\left({\rm Re}\left[\frac{|\alpha_k|}{\alpha_k}\frac{\beta_k}{|\beta_k|}\right]\right)}
\label{phase-adiabatic}\,.
\end{eqnarray}
The result of squeezing parameters is plotted in orange in Figure~{\ref{fig1}}.

\subsection{Comparison of squeezing parameters between two formalisms}
\label{3.3}

The squeezing parameters in the two formalisms that apply the instantaneous vacuum state and the adiabatic vacuum state are shown numerically in Figure~{\ref{fig1}} where 
we normalized the scale factor at the reheating time as $a(\eta_1)=1$.
The reheating time is then determined by the Hubble parameter by   $H=a^\prime(\eta))/a^2(\eta)=1/\eta_1$.
For the GUT scale inflation $H\sim 10^{14}$ GeV, the reheating time is estimated as
\begin{eqnarray}
\eta_1=10^{-14}\left(\frac{10^{14}\,{\rm GeV}}{H}\right)\,{\rm GeV}^{-1} \,.
\end{eqnarray}
Thus, the reheating time is close to the origin in Figure~{\ref{fig1}}. We consider a mode of, say, $k=10^2$ GeV at the reheating, which undergoes red shift and becomes $10^{-2}$ Hz at present, the time of horizon exit during inflation and horizon re-entry in radiation dominated phase are estimated respectively by
$\eta_{\rm exit}=-2\pi/k\sim - 0.06$ GeV$^{-1}$ and 
$\eta_{\rm entry}=2\pi/k\sim 0.06$ GeV$^{-1}$.

In the formalism that applies the instantaneous vacuum, the squeezing parameter $r_k$ in Eq.~(\ref{r-instantaneous}) keeps growing slowly in both the de Sitter and radiation dominated phases, except for a rapid increase at the instantaneous reheating.
The squeezing parameter eventually approaches approximately the same value with the result of the formalism that uses the adiabatic vacuum (See the left panel of Figure~{\ref{fig1}}). On the other hand, the asymptotic value of the effective phase Eq.~(\ref{phase-instantaneous}) of the former formalism is smaller than the result of the latter formalism (See the right panel of Figure~{\ref{fig1}}).

In the latter formalism, the squeezing parameter $r_k$ becomes constant since the Bogoliubov coefficients are calculated at the reheating time as in Eqs.~(\ref{bogoliubov1}) and (\ref{bogoliubov2}).
Hence, the particle production occurs at the reheating time, that is, when the adiabaticity of the universe is broken. The effective phase $\theta_k +\phi_k$ is also constant and given by $\pi/2$ with the above parameters. Note that in~\cite{Kanno:2021gpt}, the noise correlation of gravitons shows a maximum value with the effective phase. The finite effective phase tends to have the nonclassicality of primordial gravitational waves survive until today as shown in~\cite{Kanno:2018cuk}.

In the case of gravitons, the qualitative features of squeezing for different wavemnumbers are almost the same.
Hence, we would say that there is no significant difference in the squeezing parameter $r_k$ between the two formalisms except for the effective phase of squeezing. 

\section{Squeezed quantum states of axions}

Axions are one of the leading candidates for dark matter~\cite{Preskill:1982cy,Abbott:1982af,Dine:1982ah,Ipser:1983mw}.
Since the axion mass is generically lighter than the energy scale of inflation,
quantum fluctuations of axions during inflation are generated
and the state of axions are squeezed. In this section, we examine the squeezing process of axions by using the two formalisms and, in particular, we see the effect of the axion mass on the squeezing.

The action of the axion $\phi$ in the expanding universe is written as
\begin{eqnarray}
  S = -\frac{1}{2}\int d\eta\ a^2 \left[  (\partial_\mu \phi)^2   + m^2 \phi^2   \right] \ ,
\end{eqnarray}
where $m$ is the mass of the axion. Note that the axion potential at low energies has the form of $M^4(1-\cos{\phi/f_a})$, where $M$ is the energy scale that axions arise and $f_a$ is symmetry breaking scale.  However, if we consider only small $\phi<f_a$ displacement from the potential minimum, we can study axions in a model independent way. And the potential can be expanded as a Taylor series. Then the dominant term becomes the above mass term.

We define a new variable $x_{\bf k}(\eta)=a(\eta)\phi_{\bf k}(\eta)$ after expanding the axion field in terms of its corresponding Fourier modes. The Hamiltonian is then given by
\begin{eqnarray}
   \int d^3{\bf k}\,H_{\bf k} = \frac{1}{2}  \int d^3{\bf k}\,  \left[  \pi_{\bf k}\,\pi_{-\bf k}   + k^2 x_{\bf k}\,x_{-\bf k}   +    \frac{a'}{a}  \left(x_{\bf k}\,\pi_{-\bf k} +x_{-\bf k} \,\pi_{\bf k} \right)
  + m^2 a^2 x_{\bf k}\, x_{-\bf k} \right] \ ,
\end{eqnarray}
where $\pi_{\bf k}$ is the conjugate momentum. 
The equations of motion deduced from the Hamiltonian are
\begin{eqnarray}
 x_{\bf k}' &=& \frac{\partial H_{\bf k}}{\partial \pi_{-\bf k}}  =  \pi_{\bf k} +  \frac{a'}{a}  x_{\bf k} \,,
 \label{eom-y3}\\
  \pi_{\bf k}' &=& -\frac{\partial H_{\bf k}}{\partial x_{-\bf k}}  = -k^2  x_{\bf k}  -   \frac{a'}{a}  \pi_{\bf k}  -m^2 a^2 x_{\bf k}  
\label{eom-p3}\ . 
\end{eqnarray}

The second order differential equations for the canonical variables by using the above Eqs.~(\ref{eom-y3}) and (\ref{eom-p3}) are given by
\begin{eqnarray}
 &&x_{\bf k}^{\prime\prime} +   \left(   k^2  - \frac{a''}{a} +m^2a^2 \right) x_{\bf k}   = 0\,,
\label{eom-y4}\\
 && \pi_{\bf k}'' -\frac{2m^2a\,a^\prime}{k^2+m^2a^2}\,\pi_{\bf k}^\prime +   \left(   k^2  + m^2a^2 + a\left(\frac{a^\prime}{a^2}\right)^\prime -\frac{2m^2a^{\prime\,2}}{k^2+m^2a^2}\,  \right) \pi_{\bf k}   = 0
\label{eom-p4} \ .
\end{eqnarray}

\subsection{Squeezing in instantaneous vacuum states}
Let us first examine the squeezing parameters of the axions in the instantaneous vacuum.
The scale factor in de Sitter and radiation dominated phases is respectively given in Eq.~(\ref{history}).

\subsubsection{Evolution in de Sitter phase}

In de Sitter phase, Eq.~(\ref{eom-y4}) becomes
\begin{eqnarray}
 x_{\bf k}'' +   \left(   k^2  -    \frac{2+\frac{m^2}{H^2}}{(\eta-2\eta_1)^2}  \right) x_{\bf k}   = 0 \,, 
\end{eqnarray}
where $\eta_1$ is the reheating time. Note that the equation for $\pi_{\bf k}$ is redundant in the following discussion.
The positive frequency mode function is expressed as
\begin{eqnarray}
{\cal U}_k  (\eta )  = \frac{\sqrt{\pi}}{2}\sqrt{-(\eta-2\eta_1)}\,e^{i\frac{(2\nu+1)\pi}{4}}\,
 H_\nu^{(1)}  (-k(\eta-2\eta_1) ) \ ,
\label{positivefreq-ds}
\end{eqnarray}
where $\nu = \sqrt{9/4- m^2/H^2}$.
Note that by using the asymptotic formula of the Hankel function of the first kind:
\begin{eqnarray}
H_\nu^{(1)}  (x )  \simeq 
   \sqrt{\frac{2}{\pi x}} \exp\left( ix-\frac{2\nu+1}{2}\pi \right) \ ,
\end{eqnarray}
we can check that the mode function behaves as a positive frequency mode in Minkowski space in the remote past $-k\eta$.
The general solution of $x_{\bf k}$ can be expanded in terms of the positive and the negative frequency modes such as
\begin{eqnarray}
x_{\bf k}  (\eta) =  {\cal U}_k (\eta) {\cal A} 
+ {\cal U}_k^* (\eta) {\cal A}^\dagger  \ ,
\end{eqnarray}
where ${\cal A}$ is an arbitrary operator.
Then the annihilation operator is obtained by using the canonical variables
\begin{eqnarray}
  \hat{a}_{\bf k} (\eta) &=&
 \sqrt{ \frac{k}{2}} x_{\bf k} (\eta ) + \frac{i}{\sqrt{ 2k}} \pi_{\bf k} (\eta )\,,
 \nonumber \\
&=&\Bigl(f(\eta)+ig(\eta)\Bigr){\cal A}+\Bigl(f^*(\eta)+ig^*(\eta)\Bigr){\cal A}^\dagger\,,
\label{annihilation-ds-massive}
\end{eqnarray}
where we have defined 
\begin{eqnarray}
f(\eta)=\sqrt{\frac{k}{2}}\,{\cal U}_k(\eta)\,,\qquad g(\eta)=\frac{1}{\sqrt{2k}}\left({\cal U}_k^\prime\left(\eta\right)-\frac{a^\prime}{a}\,{\cal U}_k(\eta)\right)\,.
\end{eqnarray}
Repeating the same procedure as we did in Section~\ref{3.1.1}, the ${\cal A}$ is found to be
\begin{eqnarray}
{\cal A}=\Bigl(f^*(\eta_0)-ig^*(\eta_0)\Bigr)\hat{a}_{\bf k}(\eta_0)-\Bigl(f^*(\eta_0)+ig^*(\eta_0)\Bigr)\hat{a}^\dagger_{-\bf k} (\eta_0)\,,
\end{eqnarray}
where $\eta_0$ is the initial time. Substituting ${\cal A}$ and ${\cal A}^\dag$ back into Eq.~(\ref{annihilation-ds-massive}), we obtain the time evolution of the annihilation operator in de Sitter phase in the form,
\begin{eqnarray}
\hat{a}_{\bf k} (\eta)=\gamma^{\rm I}_k(\eta)\,\hat{a}_{\bf k} (\eta_0)+\delta^{\rm I}_k(\eta)\,\hat{a}^\dagger_{-{\bf k}}(\eta_0)\,,
\end{eqnarray}
where
\begin{eqnarray}
\gamma^{\rm I}_k(\eta)&=&\Bigl(f(\eta)+ig(\eta)\Bigr)\Bigl(f^*(\eta_0)-ig^*(\eta_0)\Bigr)-\Bigl(f^*(\eta)+ig^*(\eta)\Bigr)\Bigl(f(\eta_0)-ig(\eta_0)\Bigr)\,,\\
\delta^{\rm I}_k(\eta)&=&\Bigl(f^*(\eta)+ig^*(\eta)\Bigr)\Bigl(f(\eta_0)+ig(\eta_0)\Bigr)-\Bigl(f(\eta)+ig(\eta)\Bigr)\Bigl(f^*(\eta_0)+ig^*(\eta_0)\Bigr)\,.
\end{eqnarray}

\subsubsection{Evolution in  radiation dominant phase}

In the radiation dominant phase, Eq.~(\ref{eom-y4}) is written as
\begin{eqnarray}
x_{\bf k}'' +    \left(  k^2   + m^2 C^2 \eta^2  \right) x_{\bf k}  = 0   \ ,
\end{eqnarray}
where $C\equiv1/(H\eta_1^2)$. 
One independent solution is found to be the parabolic cylinder function,
\begin{eqnarray}
{\cal V}_k  (\eta )  = (2mC)^{-\frac{1}{4}} e^{-\frac{\pi k^2}{8mC}}
 D_{\lambda}  (z ) \ ,
 \label{positivefreq-radiation}
\end{eqnarray}
where $\lambda = -1/2- i k^2/(2mC)$ and $z=\sqrt{i2mC}\,\eta$.
Here, we used the relation between the parabolic cylinder function $D$ and a Whittaker function $W$:
\begin{eqnarray}
D_{\lambda}  (z ) = 2^{\frac{2\lambda+1}{4}} z^{-\frac{1}{2}}\,
 W_{\frac{2\lambda+1}{4},-\frac{1}{4}}  \left(\frac{z^2}{2} \right) \ .
 \label{cylinder}
\end{eqnarray}
The normalization is determined by using the formula for $|z|\rightarrow\infty$
\begin{eqnarray}
D_\lambda (z) \simeq 
z^{\lambda}e^{-\frac{z^2}{4}} \,.
\end{eqnarray}
Thus the general solutions can be written as
\begin{eqnarray}
x_{\bf k}  = 
 {\cal V}_k  (\eta )\,{\cal B} +  {\cal V}^*_k  (\eta )\, {\cal B}^\dagger\,.
\end{eqnarray}
Repeating the same procedure as in Sectin~{\ref{3.1.2}}, the ${\cal B}$ is solved as
\begin{eqnarray}
{\cal B}=\Bigl(h^*(\eta_0)-ij^*(\eta_0)\Bigr)\hat{a}_{\bf k}(\eta_1)-\Bigl(h^*(\eta_0)+ij^*(\eta_0)\Bigr)\hat{a}^\dagger_{-\bf k} (\eta_1)\,,
\end{eqnarray}
where we defined
\begin{eqnarray}
h\left(\eta\right)=\sqrt{\frac{k}{2}}\,{\cal V}_k\left(\eta\right)\,,\qquad
j\left(\eta\right)=\frac{1}{\sqrt{2k}}\left({\cal V}_k^\prime\left(\eta\right)-\frac{a^\prime}{a}\,{\cal V}_k(\eta)\right)\,.
\end{eqnarray}
Thus, the time evolution of the annihilation operator in radiation dominated phase is given by
\begin{eqnarray}
\hat{a}_{\bf k} (\eta)=\gamma^{\rm R}_k(\eta)\,\hat{a}_{\bf k} (\eta_1)+\delta^{\rm R}_k(\eta)\,\hat{a}^\dagger_{-{\bf k}}(\eta_1)\,,
\end{eqnarray}
where $\eta_1$ is the reheating time and
\begin{eqnarray}
\gamma^{\rm R}_k(\eta)&=&\Bigl(h(\eta)+ij(\eta)\Bigr)\Bigl(h^*(\eta_1)-ij^*(\eta_1)\Bigr)-\Bigl(h^*(\eta)+ij^*(\eta)\Bigr)\Bigl(h(\eta_1)-ij(\eta_1)\Bigr)\,,\\
\delta^{\rm R}_k(\eta)&=&\Bigl(h^*(\eta)+ij^*(\eta)\Bigr)\Bigl(h(\eta_1)+ij(\eta_1)\Bigr)-\Bigl(h(\eta)+ij(\eta)\Bigr)\Bigl(h^*(\eta_1)+ij^*(\eta_1)\Bigr)\,.
\end{eqnarray}

\subsubsection{Squeezing parameters}

\begin{figure}[tbp]
\vspace{-2cm}
\includegraphics[width=10cm]{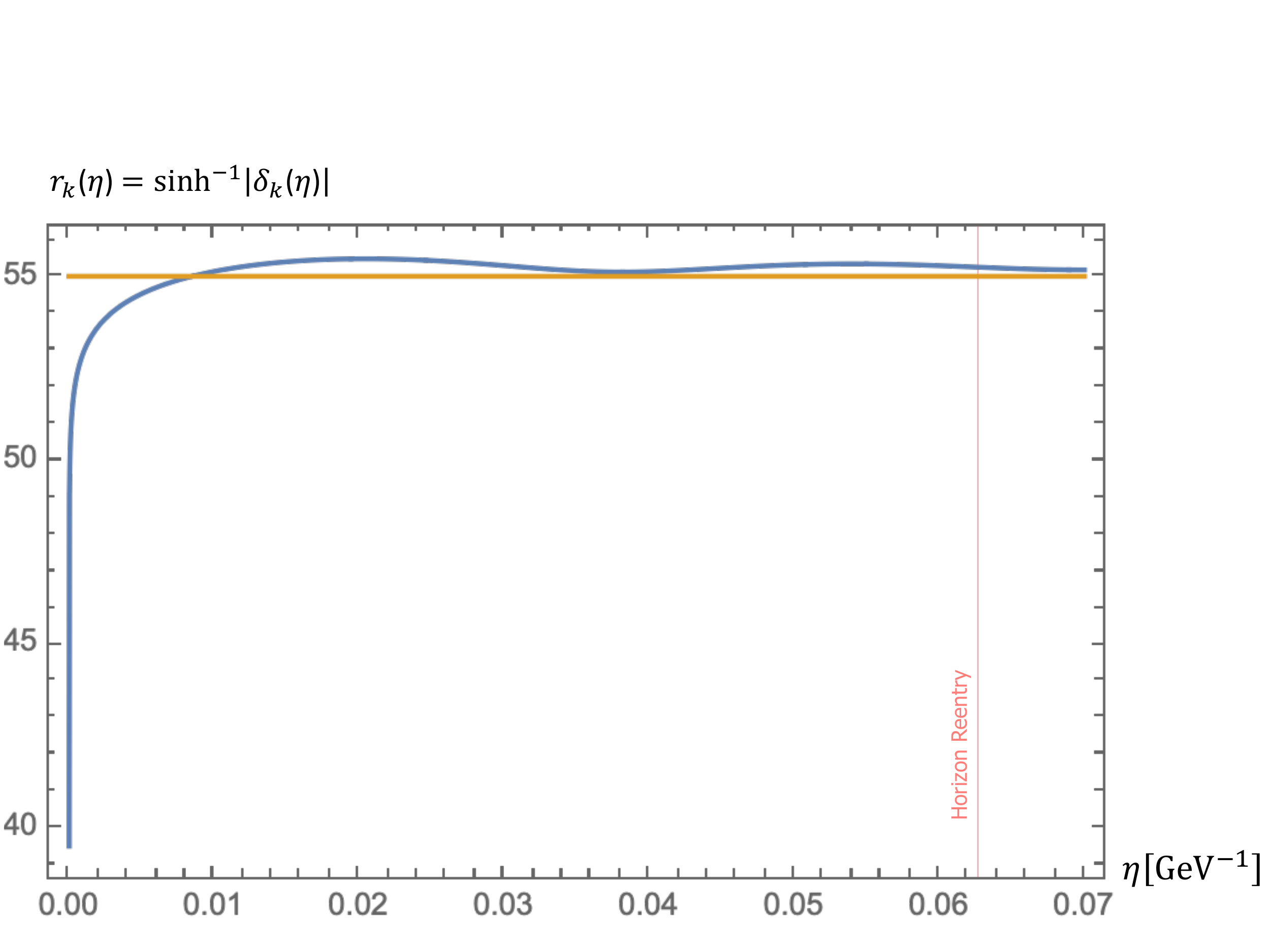}
\centering
\vspace{-0.3cm}
\caption{Plots of time evolution of the squeezing parameter $r_k$ for $k=10^2$ GeV, $\eta_1=10^{-14}\, {\rm GeV}^{-1}$ and $H=10^{14}$ GeV with the axion mass 1 $\mu$eV. The blue line is for the instantaneous vacuum, the orange is for the adiabatic vacuum. The vertical red line shows the time of horizon re-entry $2\pi/k$.} 
\label{fig2}
\end{figure}

\begin{figure}[t]
\begin{center}
\vspace{-2cm}
\hspace{-1.2cm}
\begin{minipage}{8cm}
\includegraphics[height=6.5cm]{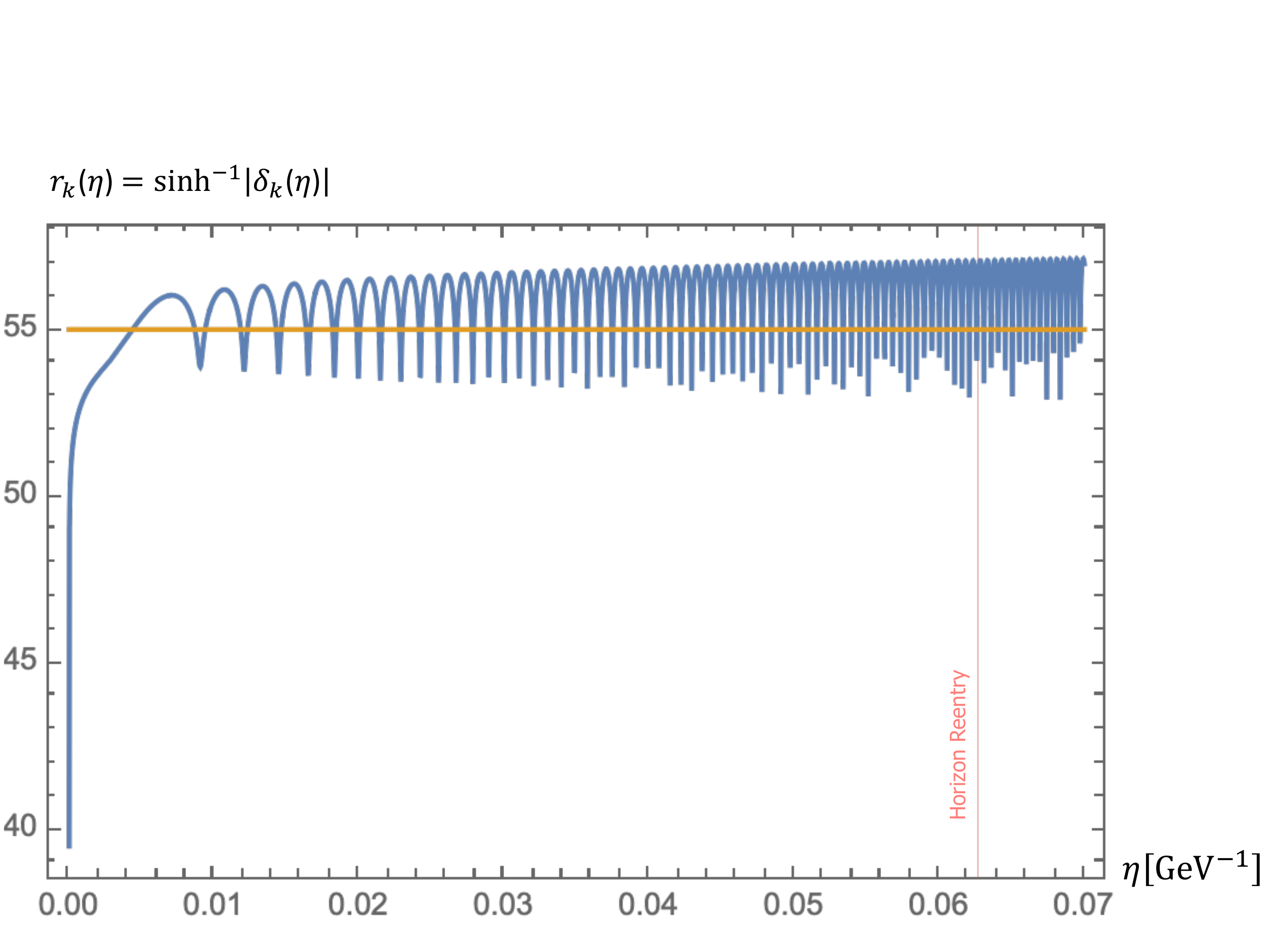}\centering
\end{minipage}
\begin{minipage}{8cm}
\hspace{0.4cm}
\vspace{0.2cm}
\includegraphics[height=6.4cm]{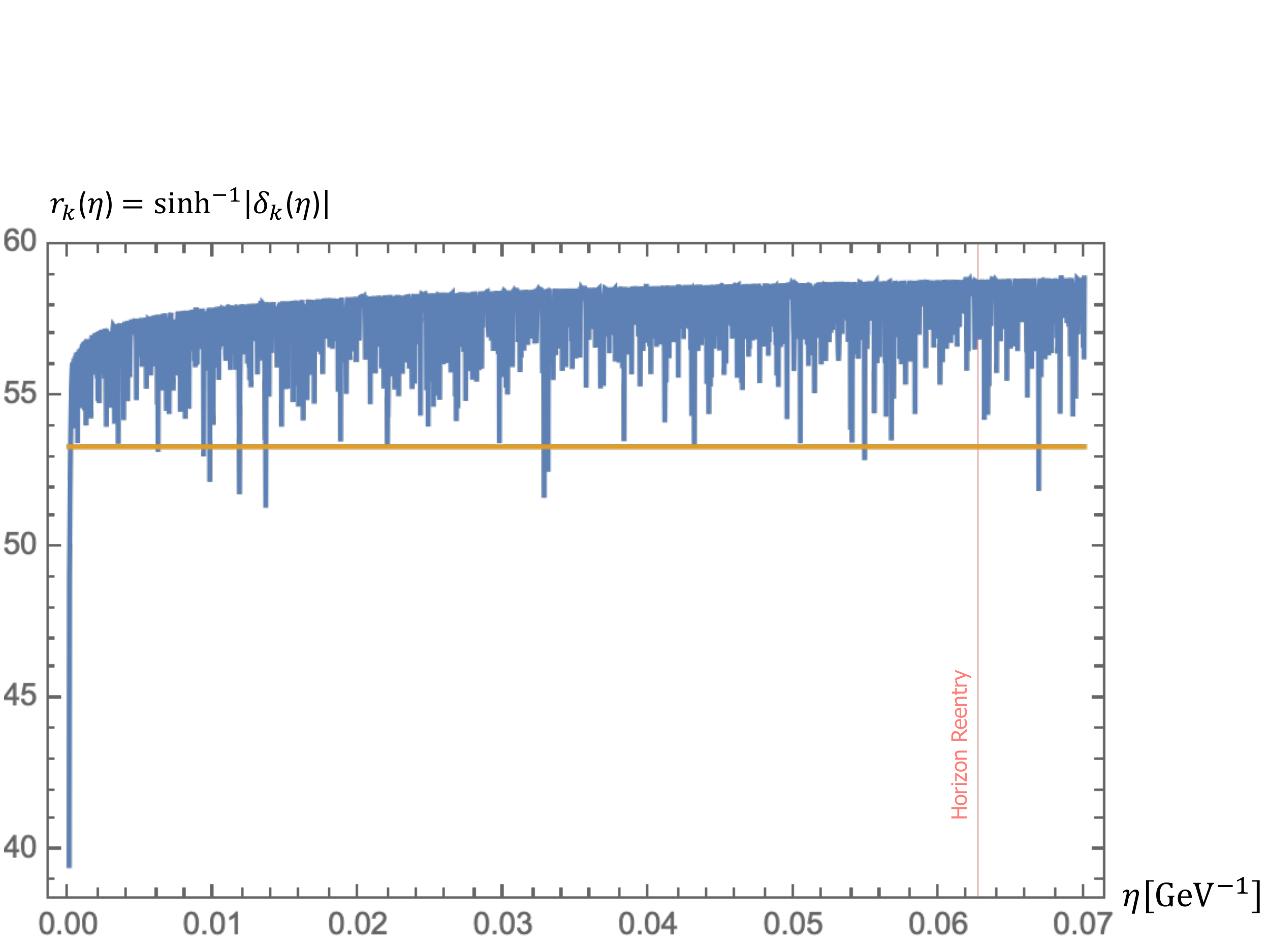}
\end{minipage}
\caption{Plots of time evolution of the squeezing parameter $r_k$ for $k=10^2$ GeV, $\eta_1=10^{-14}\, {\rm GeV}^{-1}$ and $H=10^{14}$ GeV with the axion mass 1 eV (left) and 1 keV (right). The blue line is for the instantaneous vacuum state, the orange is for the adiabatic vacuum state. The vertical red line shows the time of horizon re-entry $2\pi/k$.}
\label{fig3}
\end{center}
\end{figure}

The time evolution of the annihilation operator from the beginning of the de Sitter phase till an arbitrary time in radiation dominated phase can be expressed as
\begin{eqnarray}
\hat{a}_{\bf k}(\eta)=\gamma_k(\eta)\hat{a}_{\bf k}(\eta_0)+\delta_k(\eta)\hat{a}^\dagger_{-{\bf k}}(\eta_0)\,,
\end{eqnarray}
where
\begin{eqnarray}
\gamma_k(\eta)&=&\gamma^{\rm R}_k(\eta)\gamma^{\rm I}_k(\eta_1)+\delta^{\rm R}_k(\eta)\delta^{{\rm I}*}_k(\eta_1)\,,\\
\delta_k(\eta)&=&\gamma^{\rm R}_k(\eta)\delta^{\rm I}_k(\eta_1)+\delta^{\rm R}_k(\eta)\gamma^{{\rm I}*}_k(\eta_1)\,.
\end{eqnarray}
The squeezing parameter and the effective phase of the axion are given by
\begin{eqnarray}
r_k(\eta) &=& \sinh^{-1} |\delta_k(\eta) |\,, \\
\theta_k(\eta)+\varphi_k(\eta)&=&\frac{1}{2}\arccos{\left({\rm Re}\left[\frac{|\gamma_k(\eta)|}{\gamma_k(\eta)}\frac{\delta_k(\eta)}{|\delta_k(\eta)|}\right]\right)}\,.
\end{eqnarray}
The result of the squeezing parameters is plotted in blue in Figure~{\ref{fig2}} where the axion mass is 1 $\mu$eV. In Figure~\ref{fig3}, we plotted $r_k(\eta)$ in blue for the axion mass 1 eV (Left) and 1 keV (Right).

\subsection{Squeezing in adiabatic vacuum states}

In this subsection, we discuss the squeezing of axions in the adiabatic vacuum. In de Sitter phase, the positive frequency mode was given in Eq.~(\ref{positivefreq-ds}). On the other hand, the positive frequency mode in radiation dominated phase is defined in the remote future and given by the same form as Eq.~(\ref{positivefreq-radiation}). In fact, Eq.~(\ref{positivefreq-radiation}) for large $k$ becomes~\cite{Buchholz}
\begin{eqnarray}
 {\cal V}_k(\eta)\sim\frac{1}{\sqrt{2k}}e^{-ik\eta}\,.
\end{eqnarray}
This can be checked by using Eq.~(\ref{cylinder}) 
and then using the asymptotic expansion of Whittaker functions for large index:
\begin{eqnarray}
 W_{\kappa,\frac{\mu}{2}}(z)
 \sim 2^{-\frac{1}{2}}
 \left(\frac{z}{\kappa}\right)^{\frac{1}{4}}
 e^{\kappa\ln(\frac{\kappa}{e})+ i\left(\pi\kappa-\frac{\pi}{4}-2\sqrt{z\kappa}\right)}\,,\qquad{\rm for}\quad{\rm Im}\,\kappa\neq 0\,.
\end{eqnarray}
Then the Bogoliubov transformation between the ${\cal U}_k$ and ${\cal V}_k$ is written as
\begin{eqnarray}
  {\cal U}_k (\eta) 
  =   \gamma_k\,  {\cal V}_k (\eta  )  + \delta_k\,  {\cal V}^*_{k} (\eta) \,,
\end{eqnarray}
where the Bogoliubov coefficients are calculated at the reheating time $\eta_1$ as below
\begin{eqnarray}
  \gamma_k &=&
  i \bigl\{  {\cal V}^*_k (\eta  )  {\cal U}'_k (\eta)  -   {\cal U}_k (\eta) {\cal V}^{*\prime }_k (\eta  )   \bigr\}\big|_{\eta=\eta_1}   \nonumber\\
  &=&  i\frac{\sqrt{\pi}}{2}e^{i(2\nu+1)\frac{\pi}{4}}
  (2mC)^{-\frac{1}{4}} e^{-\frac{\pi k^2}{8mC}}
\left[  D^*_{\lambda}  (z_1 ) 
  \left\{  -\frac{\nu+ \frac{1}{2}}{\sqrt{\eta_1}}H_\nu^{(1)}  (k\eta_1 ) 
  + k\sqrt{\eta_1} H_{\nu+1}^{(1)}  (k\eta_1 ) \right\} \right.\hspace{1cm}\nonumber\\
&& \left.\hspace{5.5cm}
-  e^{-i\frac{\pi}{4}}\sqrt{2mC}\sqrt{\eta_1}  H_\nu^{(1)}  (k\eta_1 )
\left\{ \frac{1}{2}   z^*_1 D^*_{\lambda}  (z_1 ) 
     -D^*_{\lambda+1}  (z_1 )\right\} \right]
  \ ,\hspace{1cm}
\end{eqnarray}
where we defined $z_1=(-4m^2C^2)^{1/4}\eta_1$ and
\begin{eqnarray}
  \delta_k &=&
 - i \bigl\{  {\cal V}_{k} (\eta  )  {\cal U}'_k (\eta)  -   {\cal U}_k (\eta) {\cal V}^{\prime}_{k} (\eta  )   \bigr\}\big|_{\eta=\eta_1}   \nonumber\\
  &=&-i\frac{\sqrt{\pi}}{2}e^{i(2\nu+1)\frac{\pi}{4}}
  (2mC)^{-\frac{1}{4}} e^{-\frac{\pi k^2}{8mC}}
\left[  D_{\lambda}  (z_1 ) 
  \left\{  -\frac{\nu+ \frac{1}{2}}{\sqrt{\eta_1}}H_\nu^{(1)}  (k\eta_1 ) 
  + k\sqrt{\eta_1} H_{\nu+1}^{(1)}  (k\eta_1 ) \right\} \right.\hspace{1cm}\nonumber\\
&& \left. \hspace{5.5cm}
-  e^{i\frac{\pi}{4}}\sqrt{2mC}\sqrt{\eta_1}  H_\nu^{(1)}  (k\eta_1 )
\left\{ \frac{1}{2}   z_1 D_{\lambda}  (z_1 ) 
     -D_{\lambda+1}  (z_1 )\right\} \right]
  \,.\hspace{1cm}
\end{eqnarray}
Let us see the mass dependence of superhorizon modes ($k\eta_1\rightarrow 0$)
on the Bogoiubov coefficients. We can use the following approximations  
\begin{eqnarray}
H_\nu^{(1)}  (x ) \simeq
       -\frac{i}{\pi} \Gamma (\nu ) \left( \frac{x}{2} \right)^{-\nu}  \ ,\hspace{1cm} {\rm for}\quad x\ll 1
\end{eqnarray}
and
\begin{eqnarray}
D_\lambda  (x ) \simeq
2^{\frac{\lambda}{2}}
\frac{\Gamma (\frac{1}{2} )}{\Gamma (\frac{1}{2}-\frac{\lambda}{2} )}\,,
\hspace{1cm} {\rm for}\quad x\ll 1
\end{eqnarray}
where we can apply the following formula for the Gamma function
\begin{eqnarray}
 \left| \Gamma (x+iy ) \right|  \simeq
       \sqrt{2\pi} \left| y\right|^{x-\frac{1}{2}} 
       \exp\left(-\frac{\pi}{2} y\right) \,,
       \hspace{1cm} {\rm for}\quad y\gg 1\,.
\end{eqnarray}
Then we find the Bogoiubov coefficient $\left| \delta_k \right|$ expressed as
\begin{eqnarray}
 \left| \delta_k \right| 
  =\frac{\Gamma(\nu)}{\sqrt{\pi}}2^{\nu-\frac{3}{2}}\left(\nu-\frac{1}{2}\right)\frac{1}{(k\eta_1)^{\nu+\frac{1}{2}}}
  \label{bogoliubov-delta}
  \,.
\end{eqnarray}
For the massless case $\nu=3/2$, we reproduce Eq.~(\ref{bogoliubov-beta}):
\begin{eqnarray}
 \left| \delta_k \right| 
 =
  \frac{1}{2k^2\eta_1^2}=|\beta_k|
  \ .
\end{eqnarray}
The above result clearly shows the effect of mass  $\nu = \sqrt{9/4- m^2/H^2}<3/2$ reduces the magnitude of the squeezing.
Intuitively, it is reasonable result because particle creation could be hard for heavier particles. The squeezing parameters for the axion mass 1 $\mu$eV is plotted in orange in Figure~{\ref{fig2}}. Similarly, the case of heavier masses (1 eV and 1 keV) are plotted in the left and right panel respectively in Figure~\ref{fig3}.

\subsection{Comparison of squeezing parameters between two formalisms}
From the numerical result of Figure~\ref{fig2}, we see that the time evolution of the squeezing parameter of the axion is almost the same as that of the graviton in the left panel of  Figure~\ref{fig1} when the axion mass is very light (1 $\mu$eV). The asymptotic value of the squeezing of the two formalisms is also close enough. However, as the axion mass increases, there appears discrepancy in the behaviour of the time evolution of squeezing between the two formalisms as shown in Figure~\ref{fig3}. The squeezing in the instantaneous vacuum shows negative dampling oscillation while the squeezing in the adiabatic vacuum stays constant. This oscillation occurs when the Hubble friction term in the Klein-Gordon equation becomes negligible compared with the mass term when the expansion of the Universe becomes slower. The Hubble parameter when the oscillation starts is then given by $H_{\rm osc}=\eta_1 /\eta_{\rm osc}^2\simeq m$. And we find the beginning time of the oscillation is estimated as
\begin{eqnarray}
\eta_{\rm osc}
 \simeq \sqrt{\frac{\eta_1}{m}}
  \ .
\end{eqnarray}
However, we are not sure whether this oscillation suggests the particle creation. Another different point is that 
the squeezing increases as the axion mass increases from 1 eV to 1 keV in the instanteneous vacuum while the value of the squeezing in the adiabatic vacuum decreases from 55.017 to 53.3285. This decrease in squeezing in the adiabatic vacuum is consistent with the analytical result in Eq.~(\ref{bogoliubov-delta}). On the other hand, the increase in the squeezing in the instanteneous vacuum is counterintuitive because the increase of squeezing means that more particle production occurs.

%
%

\section{Cases of non-instantaneous reheating}
\label{5}

In this section, we take into account the non-instantaneous reheating phase. The reheating phase occurs after slow-roll conditions for inflaton are violated and the time-averaged equation of state becomes the same as that of cold dark matter. Thus the inflaton provides the classical background matter in the reheating phase.

Then, the scale factor evolves as the following de Sitter, reheating and radiation dominated phases,
\begin{eqnarray}
 a (\eta )=\left\{
\begin{array}{l}
\vspace{0.2cm}
 \frac{1}{-H  (  \eta -2 \eta_1 )}  \hspace{2.4cm}   {\rm for}  \quad   \eta   <  \eta_1\,,\\
 \vspace{0.2cm}
  \frac{(\eta+\eta_1)^2}{4H\eta_1^3 }  \hspace{2.9cm}   {\rm for}  
  \quad   \eta_1 < \eta   <  \eta_2\,,\\
\frac{\eta_1+\eta_2}{2H\eta^3_1}\left(\eta-\frac{\eta_2 -\eta_1}{2}\right) \hspace{1.0cm}     {\rm for} \quad     \eta_2  <  \eta   \ ,
\end{array}
\right.             
\label{new_history}
\end{eqnarray}
where the scale factor is smoothly connected up to the first order of derivative at $\eta_1$ and $\eta_2$ respectively. For simplicity, we consider gravitons and the equation of motion of the graviton is given in Eq.~(\ref{eom-y2}). The positive frequency mode in de Sitter phase is already obtained in Eq.~(\ref{uk}). The equation of motion in the reheating phase is expressed as
\begin{eqnarray}
\hat{y}_{\bf k}'' +   \left(   k^2  -    \frac{2}{(\eta+\eta_1)^2}  \right) \hat{y}_{\bf k}   = 0   \,.
\end{eqnarray}
The positive fequency mode is found to be
\begin{eqnarray}
w_k  =   \frac{1}{\sqrt{ 2k}}\left(1-\frac{i}{k(\eta+\eta_1)}  \right) e^{  - ik (\eta+\eta_1)} \ .
\end{eqnarray}
The equation of motion in the radiation dominated phase is in Eq.~(\ref{eom-y-rad}). And the positive frequency mode is written as
\begin{eqnarray}
v_k  =   \frac{1}{\sqrt{ 2k}}e^{  - ik (\eta-\frac{\eta_2 -\eta_1}{2})} \ .
\end{eqnarray}
The Bogoliubov coefficients $\alpha_{1k}$, $\beta_{1k}$ between the de Sitter and the reheating phases can be obtained by the continuity relations at $\eta_1$
\begin{eqnarray}
  u_k (\eta_1) 
  &=& \alpha_{1k}\,w_k (\eta_1  )  + \beta_{1k}\, w^*_{-k} (\eta_1)  \ ,  \label{reheating-initial} \\
   u'_k (\eta_1) 
  &=&   \alpha_{1k}\, w'_k (\eta_1  )  + \beta_{1k}\,  w^{*\prime}_{-k} (\eta_1) \ .
\end{eqnarray}
Simililary, the Bogoiubov coefficients $\alpha_{2k}$, $\beta_{2k}$ between the reheating and the radiation dominated phases are given at $\eta_2$
\begin{eqnarray}
  w_k (\eta_2) 
  &=&   \alpha_{2k}\,v_k (\eta_2  )  + \beta_{2k}\,v^*_{-k} (\eta_2)  \ , \label{reheating-final}  \\
   w'_k (\eta_2) 
  &=&   \alpha_{2k}\,  v'_k (\eta_2  )  + \beta_{2k}\,  v^{*\prime}_{-k} (\eta_2) \ .
\end{eqnarray}
\begin{figure}[t]
\begin{center}
\vspace{-2cm}
\hspace{-1.2cm}
\begin{minipage}{8cm}
\includegraphics[height=6.5cm]{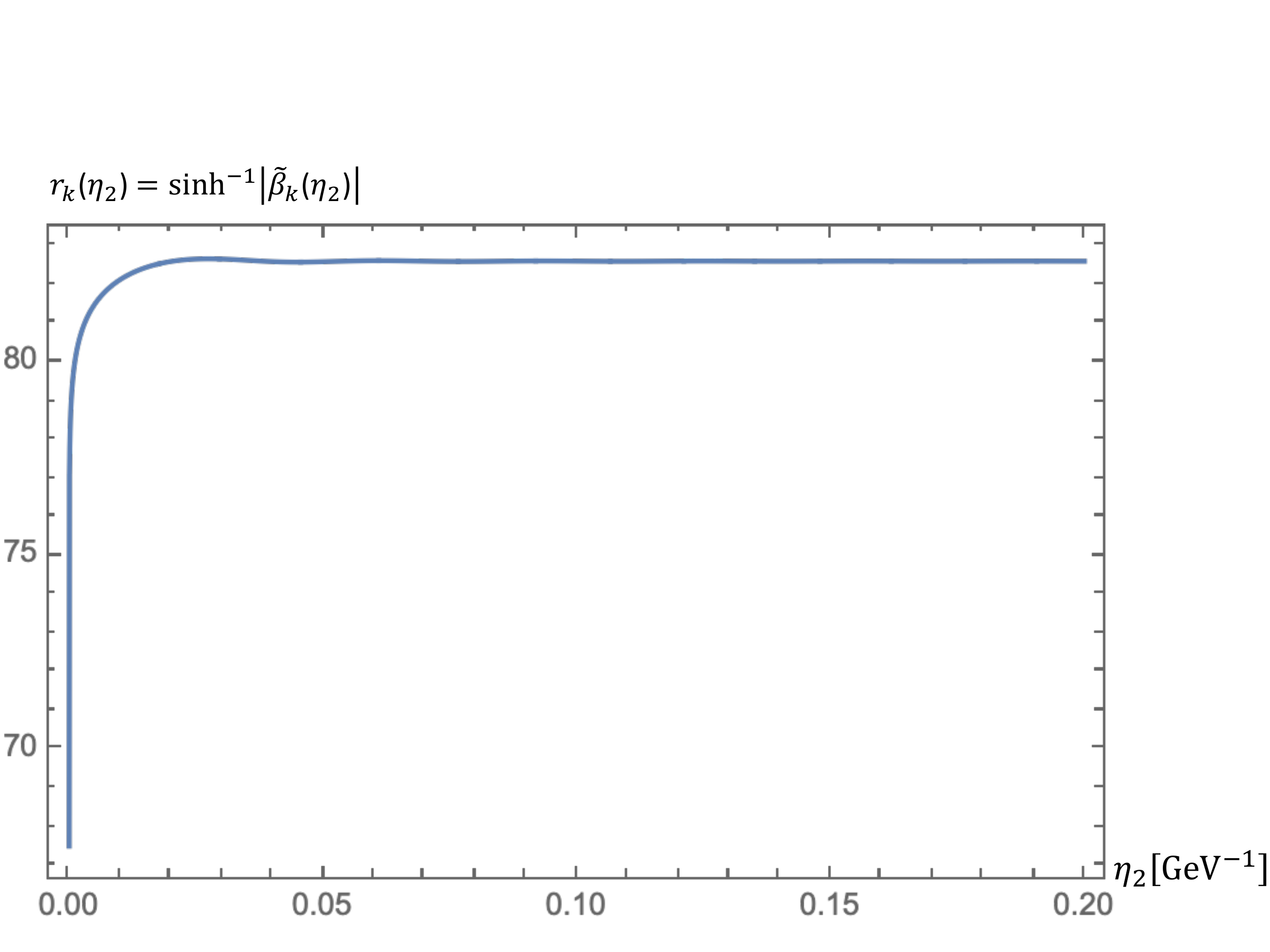}
\centering
\end{minipage}
\begin{minipage}{8cm}
\hspace{0.4cm}
\vspace{0.2cm}
\includegraphics[height=6.4cm]{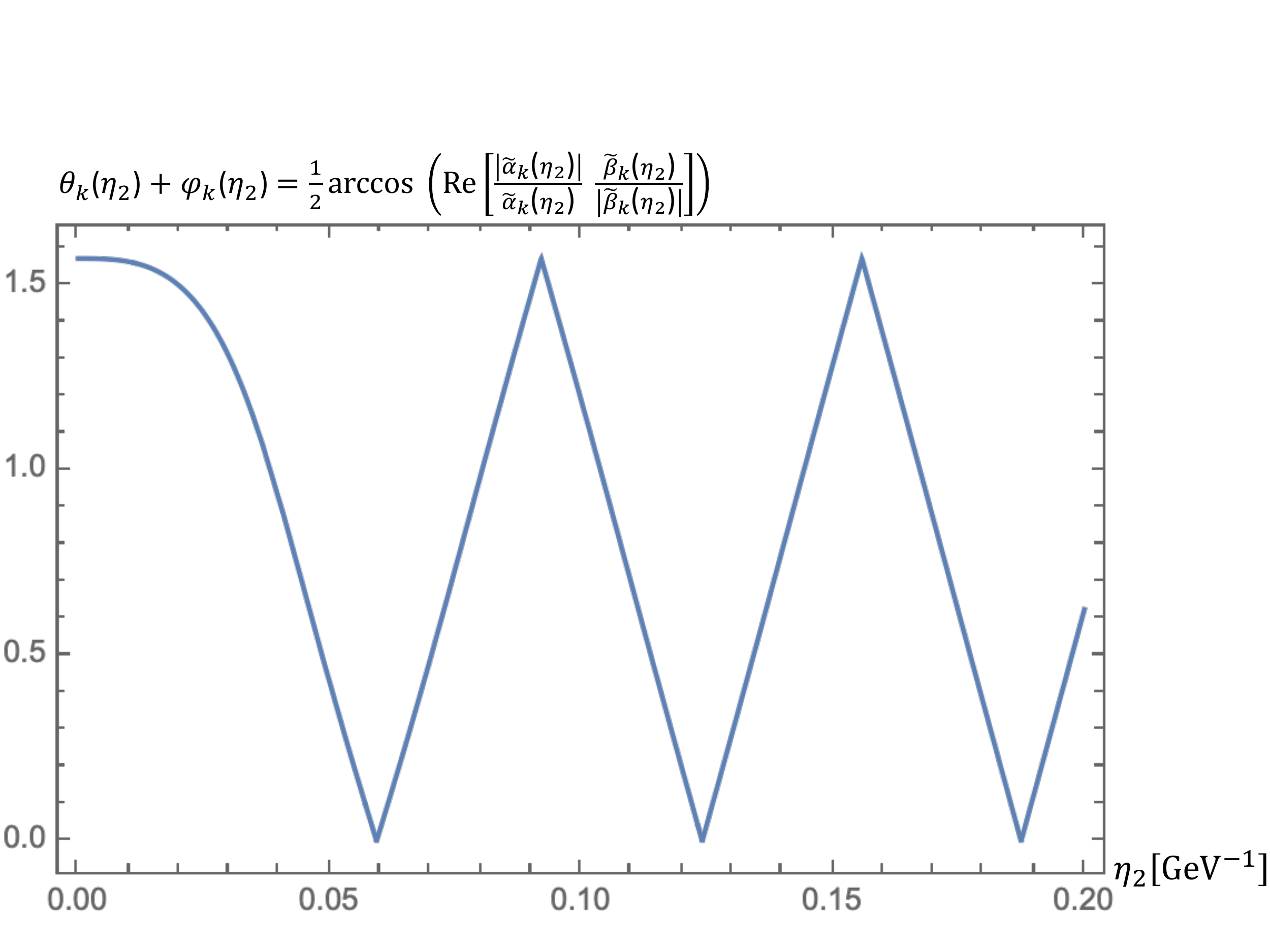}
\end{minipage}
\caption{Plots of the squeezing parameter $r_k$ (left)  and the effective phase $\theta +\varphi$ (right)
as a function of $\eta_2$ for $k=10^2$ GeV, $\eta_1=10^{-14}$ and ${\rm GeV}^{-1}$.}
\label{fig4}
\end{center}
\end{figure}
By solving the above continuity relations, the Bogoliubov coefficients are turns out to be
\begin{eqnarray}
  \alpha_{1k}    &=&  
  \left(1 - \frac{9}{8k^2 \eta_1^2}+\frac{3i}{2k  \eta_1}  - \frac{3i}{8 k^3  \eta_1^3} \right) 
  e^{3 i k  \eta_1  }\,,      \\
  \beta_{1k}   &=&  \frac{3}{8k^2\eta_1^2}
  \left( 1-\frac{i}{k\eta_1}\right)e^{- i k  \eta_1  } \ .  
\end{eqnarray}
and
\begin{eqnarray}
  \alpha_{2k}    &=&  
  \left(1-\frac{i}{k \left( \eta_1+\eta_2 \right)}  - \frac{1}{2 k^2  \left(\eta_1 +\eta_2\right)^2} \right)   
  e^{- i k  \frac{\eta_1 +\eta_2}{2}  }\,,      \\
  \beta_{2k}   &=&  
  \frac{1}{2k^2\left( \eta_1+\eta_2\right)^2}  
  e^{- 3i k  \frac{\left( \eta_1 +\eta_2\right) }{2}  } 
  \ .  
\end{eqnarray}
Notice that the above Bogoliubov coefficients depend on the time $\eta_1$ and $\eta_2$.
In order to see the effect of the reheating phase, let us consider the Bogoliubov transformation at an arbitrary time. The Bogoliubov transformation between de Sitter and the reheating phases and that between the reheating and the radiation dominated phases are respectively written as,
\begin{eqnarray}
u_k(\eta)&=&\alpha_{1k}\,w_k (\eta)  + \beta_{1k}\,  w^*_{-k} (\eta)\,,\label{relation1}\\
w_k(\eta)&=&\alpha_{2k}\,v_k (\eta)  + \beta_{2k}\,  v^*_{-k} (\eta)\,.\label{relation2}
\end{eqnarray}
Plugging Eq.~(\ref{relation2}) into Eq.~(\ref{relation1}), we find that he Bogoliubov transformation from the initial time in de Sitter till the radiation dominated phase becomes
\begin{eqnarray}
  u_k (\eta) 
  &=&   \tilde{\alpha}_k\,  v_k (\eta  )  + \tilde{\beta}_k\,  v^*_{-k} (\eta)  \ .
\end{eqnarray}
where
\begin{eqnarray}
 \tilde{\alpha}_k    &=&  \alpha_{1k}\, \alpha_{2k} + \beta_{1k}\, \beta_{2k}^*     \\
  \tilde{\beta}_k   &=&  \alpha_{1k}\, \beta_{2k} + \beta_{1k}\, \alpha_{2k}^*     \ .  
\end{eqnarray}
Note that $\tilde{\alpha}_k$ and $\tilde{\beta}_k$ depend on the time $\eta_1$ and $\eta_2$. 
If we fix the $\eta_1$, say, $\eta_1=10^{-14}\,{\rm GeV}^{-1}$, the dependence of the end of reheating time $\eta_2$ on the squeezing parameter $r_k(\eta_2)$ and the effective phase $\theta_k(\eta_2)+\varphi_k(\eta_2)$ are written as
\begin{eqnarray}
r_k(\eta_2) &=& \sinh^{-1} |\tilde{\beta}_k(\eta_2) |
\label{r-adiabatic}\,, \\
\theta_k(\eta_2)+\varphi_k(\eta_2)&=&\frac{1}{2}\arccos{\left({\rm Re}\left[\frac{|\tilde{\alpha}_k(\eta_2)|}{\tilde{\alpha}_k(\eta_2)}\frac{\tilde{\beta}_k(\eta_2)}{|\tilde{\beta}_k(\eta_2)|}\right]\right)}
\,.
\end{eqnarray}
We plotted the squeezing parameter and the effective phase as a function of $\eta_2$ in Figure~\ref{fig4}. Comparing with Figure~\ref{fig1}, we see that the squeezing is enhanced to the value 82.5967 from 55.262. The effective phase is found to oscillate and becomes zero periodically.

\section{Nonclassicality of gravitons and axions}
\label{6}

In this section, we demonstrate that the nonclassicality  of gravitons and axions are enhanced in the squeezed states. 

Let us consider a Fourier field operator $\hat{\psi}_{\bf k}(\eta)$ expanded in terms of two kinds of positive and negative frequency modes $({\bar u}_k, {\bar u}^*_k)$ and $({\bar v}_k, {\bar v}^*_k)$ respectively such as
\begin{eqnarray}
 \hat{\psi}_{\bf k}(\eta)=\hat{a}_{\bf k}\,{\bar u}_k(\eta)+\hat{a}^\dag_{-\bf k}\,{\bar u}_k^*(\eta)
 =\hat{b}_{\bf k}\,{\bar v}_k(\eta)+\hat{b}^\dag_{-\bf k}\,{\bar v}_k^*(\eta)\,,
\end{eqnarray}
where $[{\hat a}_{\bf k},{\hat a}^\dag_{-\bf p}]=\delta_{{\bf k}, {-\bf p}}$ and $[{\hat b}_{\bf k},{\hat b}^\dag_{-\bf p}]=\delta_{{\bf k}, {-\bf p}}$ should be satisfied. We can define two different vacua $\hat{a}_{\bf k}|\bar{0}_{\bf k}\rangle=0$ and $\hat{b}_{\bf k}|0_{\bf k}\rangle=0$. Then the operators $(\hat{a}_{\bf k}, \hat{a}^\dag_{-\bf k})$ and $(\hat{b}_{\bf k}, \hat{b}^\dag_{-\bf k})$ are related by the Bogolubov transformation by using the squeezing parameters
\begin{eqnarray}
\hat{a}_{\bf k} = e^{-i\theta_k}\cosh r_{k}\,  \hat{b}_{\bf k}
+ e^{i\theta_k+2i\varphi_{k}} \sinh r_{k}\,  \hat{b}^\dagger_{-\bf k}\,.
\end{eqnarray}
Combining $\hat{a}_{\bf k}|\bar{0}_{\bf k}\rangle=0$ with the above relation, we have
\begin{eqnarray}
\left(\cosh r_{k}\,\hat{b}_{\bf k}
+ e^{i\left(\theta_k+\varphi_{k}\right)} \sinh r_{k}\, \hat{b}^\dagger_{-\bf k}\right)|\bar{0}_{\bf k}\rangle=0\,.
\end{eqnarray}
If we use the relation $[{\hat b}_{\bf k},{\hat b}^\dag_{-\bf p}]=\delta_{{\bf k}, {-\bf p}}$, the above relation becomes a differential equation and the solution is written by
\begin{eqnarray}
|\bar{0}_{\bf k}\rangle 
= N_k\exp\left[e^{i\left(\theta_k+\varphi_{k}\right)}\tanh r_{k} 
\  \hat{b}^\dagger_{\bf k} \, \hat{b}^\dagger_{-\bf k}\right]
|0_{\bf k}\rangle\otimes |0_{-{\bf k}} \rangle  \ ,
\end{eqnarray}
where $N_k$ is the normalization constant.
If we expand the exponential function in Taylor series, we find
\begin{eqnarray}
|\bar{0}_{\bf k}\rangle  =  N_k \left(\, |0_{\bf k}\rangle\otimes |0_{-{\bf k}}\rangle  + e^{i\left(\theta_k+\varphi_{k}\right)} \tanh r_k\,|1_{\bf k}\rangle\otimes |1_{-{\bf k}}\rangle 
+\cdots +e^{in\left(\theta_k+\varphi_{k}\right)}\tanh^n r_k\,|n_{\bf k}\rangle\otimes |n_{-{\bf k}}\rangle\right) \,. 
\label{two-mode}
\end{eqnarray}
This is a two-mode squeezed state which consists of an infinite number of entangled particles. In particular, in the highly squeezing limit $r_k\rightarrow\infty$, the state becomes the maximally entangled state and the nonclassicality is enhanced. In cosmology, $|\bar{0}_{\bf k}\rangle$ and $|0_{\bf k}\rangle$ correspond to the Bunch-Davies vacuum and the vacuum in radiation dominated phase. We discuss the possibility to observe the gravitons and axions in this two-mode squeezed state in the next section.

\subsection{Bell inequality}
Bell inequality is a measure of entanglement and the inequality is violated if quantum correlations exist.

Let us consider two sets of non-commuting operators $A$, $A'$ and $B$, $B'$. Those operators correspond to measuring the spin along various axes and have eigenvalues $\pm 1$. They are expressed by the Pauli matrices $\sigma^i$ and unit vectors $n^i$ such as $A=n^i\sigma^i$. 
The Bell operator ${\cal B}$ is defined as~\cite{Clauser:1969ny}
\begin{eqnarray}
{\cal B}=\frac{1}{2} \left( A\otimes B+A'\otimes B+A\otimes B'-A'\otimes B'  \right)\,,
\label{bell1}
\end{eqnarray}
where the variables $A$, $A'$ and $B$, $B'$ are represented by Hermitian operators which act on the Hilbert spaces ${\cal H}_A$ and ${\cal H}_B$ respectively. If we rewrite it as a factorized form
\begin{eqnarray}
{\cal B}=\frac{1}{2}A\otimes \left(B+B'\right)+\frac{1}{2}A'\otimes\left(B-B'\right)\,,
\label{bell2}
\end{eqnarray}
then we see that the first (second) term becomes $\pm 1$ while the second (first) one vanishes because we can have either $B=B'$ or $B=-B'$. 
In classical theories, the expectation value of ${\cal B}$ then gives 
\begin{eqnarray}
|\langle{\cal B}\rangle|\leq 1. 
\label{bell}
\end{eqnarray}
This is the Bell inequality. In quantum mechanics, however, $A$ and $A'$ cannot be factored out as done in Eq.~(\ref{bell2}).
So insted, we consider the squared form of the Bell operator
\begin{eqnarray}
{\cal B}^2=I- \frac{1}{4}\left[A,A'\right]\otimes\left[B,B'\right]\,,
\end{eqnarray}
where we used the fact that the square of each operator is one, $A^2=I$, $A'^2=I$, etc and $I$ is the identity operator. Since the commutators of the Pauli matrices are non-zero\footnote{The Pauli matrices satisfy $[\sigma_a,\sigma_b]=2i\varepsilon_{abc}\,\sigma_c$ where $\varepsilon_{abc}$ is antisymmetric tensors. For local classical hidden variable theories, the commutators are zero and $\langle{\cal B}^2\rangle\leq 1$ or $|\langle{\cal B}\rangle|\leq 1$.} and each gives $2i$, we find that 
\begin{eqnarray}
\langle{\cal B}^2\rangle\leq 2\quad{\rm or}\quad |\langle{\cal B}\rangle|\leq \sqrt{2}. 
\end{eqnarray}
We see that the Bell inequality Eq.~(\ref{bell}) is violated and the maximal violation of Bell inequality in quantum mechanics becomes $\sqrt{2}$ in the case of a pair of spins.

In the case of quantum field theory, we consider the entanglement between ${\bf k}$ and $-{\bf k}$ as seen in Eq.~(\ref{two-mode}). We need to define a pseudo-spin operator
for each mode with wavenumber vector ${\bf k}$ and use two sets of non-commuting pseudo-spin operators as demonstrated in Eq.~(\ref{bell1}). Since the two sets of non-commuting pseudo-spin operators are written by the inner product with a unit vector respectively, for an appropriate configuration of angles of the inner products~\cite{Kanno:2017dci}, we get the maximal violation for each entanglement between ${\bf k}$ and $-{\bf k}$ of the Bunch-Davies vacuum $|\bar{0}_{\bf k}\rangle$:
\begin{eqnarray}
\langle {\bar 0}_{\bf k}|{\cal B}_{\bf k}|{\bar 0}_{\bf k}\rangle
=\sqrt{1+\tanh^2 2r_k}\,\xrightarrow{r_k\rightarrow\infty}\,\sqrt{2} \,,
\end{eqnarray}
where we took the infinite squeezing limit.
If we consider all modes in the Bunch-Davies vacuum, we obtain the squeezed state $|\zeta\rangle$ that consists of an infinite number of the modes of $|\bar{0}_{\bf k}\rangle$ expressed as
\begin{eqnarray}
|\zeta\rangle
= \prod_{\bf k}|\bar{0}_{\bf k}\rangle\,.
\end{eqnarray}
Then the Bell operator in the squeezed state
is given by the sum of the Bell operator for each mode and the violation of the Bell inequality becomes
\begin{eqnarray}
\langle \zeta |\sum_{\bf k} {\cal B}_{\bf k}|\zeta\rangle
\simeq\sqrt{2}\times {\rm the \ number\ of\  modes \ {\bf k}} \,. 
\end{eqnarray}
We see that the violation increases as the number of modes increases.

\subsection{Noise of gravitons and axions}
Recently, some ideas for detecting gravitons indirectly are proposed. One is to detect the amplitude of the noise of gravitons on the gravitational interferometers~\cite{Parikh:2020nrd,Kanno:2020usf,Parikh:2020kfh,Parikh:2020fhy} and another is to detect the decoherence time of an object caused by the noise of gravitons~\cite{Kanno:2020usf}. The effect of the noise comes in the form of correlation functions of gravitons in either idea. Thus, even if we consider the noise of axions instead of gravitons, the noise correlation is essential.

For instance, in~\cite{Kanno:2020usf}, the decoherence functional is written as
\begin{eqnarray}
\Gamma (\tf) \approx\frac{m^2}{8}\int^{\tf}_0\mathrm{d}t\,\Delta(\xi^i\xi^j)(t)\int^{\tf}_0\mathrm{d}t'\,\Delta(\xi^k\xi^\ell)(t')   \Big\langle \zeta\Big|\left\{\hat h_{ij}(t),\,\hat h_{k\ell}(t')\right\}\Big|\zeta\Big\rangle\,,
\label{rate}
\end{eqnarray}
where $m$ is the mass of an object, $t_{\rm f}$ is the time of decoherence and $\xi^i$ is geodesic deviation between two objects. $\Delta(\xi^i\xi^j)(t)\equiv\xi^i_1(t)\xi^j_1(t)-\xi^i_2(t)\xi^j_2(t)$ denotes a difference of $\xi^i(t)\xi^j(t)$ in the superposition.  This expression shows that one can compute the decoherence rate due to gravitons once the anti-commutator correlation function of $\hat{h}_{ij}(t)$ is given.
In the case of gravitons, we have the noise correlation of the form
\begin{eqnarray}
 \Big\langle \zeta\Big|\left\{\hat h_{ij}(t),\,\hat h_{k\ell}(t')\right\}\Big|\zeta\Big\rangle 
  = \left( \delta_{ik} \delta_{j\ell} +  \delta_{i\ell} \delta_{jk} 
                 - \frac{2}{3}  \delta_{ij}  \delta_{k\ell}      \right)  \frac{F(\Omega_m(t-t'))}{10\pi^2M_{\rm p}^2}   \,, \quad
\label{eq:noiseamp3}
\end{eqnarray}
where $M_{\rm p}$ and $\Omega_m$ are the Planck mass and cutoff frequency. We defined the anticommutator symbol $\{\cdot\,,\cdot\}$ as $\{\hat X,\hat Y\}\equiv(\hat X\hat Y+\hat Y\hat X)/2$ and 
\begin{equation}
F(\Omega_m(t-t'))\equiv \int^{\Omega_m}_0\mathrm{d}k\,k^6\mathrm{Re}\left[u^{\rm sq}_k(t){u^{\rm sq}_k}^*(t')\right] \ .\label{Fintegral}
\end{equation}
Here $u^{\rm sq}_k$ is the mode function in the squeezed state and which behaves as
\begin{equation}
u^{\rm sq}_k(t) = \frac{1}{\sqrt{2k}}e^{-i\omega t}\cosh r_k
+\frac{1}{\sqrt{2k}}e^{i\omega t} \sinh r_k
\propto e^{r_k}
\ .\label{squeez}
\end{equation}
Note that the noise correlation in the Minkowski vacuum is obtained in the limit of $r_k\rightarrow 0$. In that case, we see that the noise correlation is negligibly small and in fact the decoherence hardly occurs within the life time of the universe. 
Remarkably, the noise correlation is found to be exponentially enhanced in the squeezed state~\cite{Parikh:2020nrd,Kanno:2020usf,Parikh:2020kfh,Parikh:2020fhy}. This is because the squeezed state consists of an infinite number of entangled particles. 

\section{Conclusion}

We studied cosmological squeezing process of gravitons and axions and demonstrated that the squeezed states enhance their nonclassicality of observables.
As a cosmological model, we considered
de Sitter phase followed by instantaneous reheating leading to radiation dominated phase. We also considered the effects of reheating phase lastly in Section~\ref{5}.
The squeezing process is related to the process of particle creation. The squeezing parameters are obtained by calculating the Bogoliubov transformation between the two phases. There are two methods for calculating the Bogoliubov transformation. One is to employ instantaneous vacuum states, and the other is to adopt adiabatic vacuum states. The biggest difference between the two methods is the definition of the vacuum state. The instantaneous vacuum state is defined at every moment between the two phases while the adiabatic vacuum state is defined in the remote past during inflation phase and in the remote future in radiation dominated phase.

We analytically calculated the squeezing parameters of gravitons and axions with the two methods and then compared the results. In the case of gravitons, that is, massless particles, there appeared no qualitative difference in the squeezing parameters between the two methods. However, in the case of axions, that is, massive particles, there appeared qualitative differences. It was found that the squeezing in the instantaneous vacuum increases as the mass increases while the squeezing decreases in the adiabatic vacuum as the mass increases. Another qualitative difference was that the squeezing parameter showed a behaviour of negative damping oscillation in the instanteneous vacuum while the squeezing parameter in the adiabatic vacuum keeps constant. Such a oscillating behavior in the instantaneous vacuum appeared when the expansion of the universe becomes slower after inflation. But it is hard to interpret the behaviour as particle creation. Thus, we conclude that the adiabatic vacuum would be appropriate for discussing the particle creation. 

We then examined the effect of the reheating phase between the de Sitter and radiation dominated phases on the squeezing parameters of gravitons by using the adiabatic vacuum. It tuned out that the squeezing is enhanced due to the reheating phase. This means that particle creation increases by considering the reheating phase. The effective phase showed qualitatively different behaviour from the instantaneous reheating. The phase started to oscillate and becomes zero periodically.

In the last section, we discussed implications of squeezed states by using some measures of nonclassicality. we demonstrated that the violation of the Bell inequalities  and the noises of gravitons and axions are enhanced in squeezed states. These results would be useful for revealing the quantum nature of the origin of the universe.

\section*{Acknowledgments}
S.\,K. was supported by the Japan Society for the Promotion of Science (JSPS) KAKENHI Grant Number JP18H05862.
J.\,S. was in part supported by JSPS KAKENHI Grant Numbers JP17H02894, JP17K18778, JP20H01902.

\end{document}